\documentclass[twocolumn, trackchanges]{aastex631}
\usepackage{array}

\usepackage{graphicx}
\usepackage{amsmath}
\usepackage{float}
\usepackage{appendix}
\usepackage{macros}
\usepackage{chngcntr}

\newcommand{\ucdavis}{Department of Physics \& Astronomy, University of California, Davis, CA 95616, USA}

\begin{document}

\title{Spatial Variations of Stellar Elemental Abundances in FIRE Simulations of Milky Way-Mass Galaxies: Patterns Today Mostly Reflect Those at Formation}

\shorttitle{Spatial Variations of Metals in FIRE}
\shortauthors{Graf et al.}

\correspondingauthor{Russell L. Graf}
\email{rlgraf@ucdavis.edu}

\author[0009-0009-8310-8992]{Russell L. Graf}
\affiliation{\ucdavis}

\author[0000-0003-0603-8942]{Andrew Wetzel}
\affiliation{\ucdavis}

\author[0000-0002-5663-207X]{Matthew A. Bellardini}
\affiliation{\ucdavis}

\author[0000-0001-6380-010X]{Jeremy Bailin}
\affiliation{Department of Physics \& Astronomy, University of Alabama, Box 870324, Tuscaloosa, AL 35487-0324, USA}

\begin{abstract}
Spatial patterns of stellar elemental abundances encode rich information about a galaxy's formation history.
We analyze the radial, vertical, and azimuthal variations of metals in stars, both today and at formation, in the FIRE-2 cosmological simulations of Milky Way (MW)-mass galaxies, and we compare with the MW.
The radial gradient today is steeper (more negative) for younger stars, which agrees with the MW, although radial gradients are shallower in FIRE-2.
Importantly, this age dependence was present already at birth: radial gradients today are only modestly ($\lesssim0.01$ dex kpc$^{-1}$) shallower than at birth.
Disk vertical settling gives rise to negative vertical gradients across all stars, but vertical gradients of mono-age stellar populations are weak.
Similar to the MW, vertical gradients in FIRE-2 are shallower at larger radii, but they are overall shallower in FIRE-2.
This vertical dependence was present already at birth: vertical gradients today are only modestly ($\lesssim0.1$ dex kpc$^{-1}$) shallower than at birth.
Azimuthal scatter is nearly constant with radius, and it is nearly constant with age $\lesssim8$ Gyr ago, but increases for older stars.
Azimuthal scatter is slightly larger ($\lesssim0.04$ dex) today than at formation.
Galaxies with larger azimuthal scatter have a stronger radial gradient, implying that azimuthal scatter today arises primarily from radial redistribution of gas and stars.
Overall, spatial variations of stellar metallicities show only modest differences between formation and today; spatial variations today primarily reflect the conditions of stars at birth, with spatial redistribution of stars after birth contributing secondarily.
\end{abstract}


\section{Introduction}

The origin and evolution of elemental abundances encode vital information about a galaxy's formation history \citep{Tinsley1980}. Through stellar nucleosynthesis \citep{B2FH1957} and supernovae \citep{Baade1934}, heavy elements (beyond H and He) are ejected back into the interstellar medium (ISM), from which successive generations of more metal-rich stars are born.
Thus, the ISM and the stars that recently formed out of it reflect the current elemental abundance patterns within a galaxy.
Once formed, stars essentially retain the abundance patterns they were born with, so a galaxy's stellar population encodes its entire enrichment, and thus formation, history.

Importantly, elemental abundances encode spatial information. \citet{Searle1971} first identified negative radial gradients in gas-phase metallicity, such that smaller radii within galaxies are more enriched.
More recently, many works have quantified spatial variations of metals in gas in the Milky Way (MW), M31, and nearby galaxies \citep[for example,][]{Sanders2012, SanchezMenguiano2016, Belfiore2017, Sakhibov2018, Kreckel2019, Kreckel2020, Wenger2019, Molla2019, Zinchenko2019, Hernandez2021}. The spatial patterns of metallicity in stars of different ages today encode the history of a galaxy's ISM, subject to the complication that stars can move from their birth radii and vertical height in the disk via radial redistribution and vertical heating \citep[for example,][]{SellwoodBinney2002, Haywood2008}.

Many observational surveys have advanced our ability to measure stellar metallicity variations across the MW, such as GALAH \citep{Buder2018}, Gaia-ESO \citep{Gilmore2012}, LAMOST \citep{Cui2012}, and SDSS-APOGEE \citep{Jonsson2020}, and soon SDSS-V \citep{Kollmeier2017}, 4-m Multi-Object Spectrograph Telescope \citep[4MOST;][]{deJong2019}, the WHT Enhanced Area Velocity Explorer \citep[WEAVE;][]{Dalton2012}, and the MaunaKea Spectroscopic Explorer \citep[MSE;][]{MSE2019} will measure spectroscopy and precise elemental abundances for tens of millions of stars.
Many analyses of these surveys identified radial and vertical gradients in stars and used them to infer the spatial and temporal evolution of enrichment in the MW \citep[for example,][]{Vickers2021, Lian2022, Lian2023, Willet2023, Anders2023, Imig2023}.
These works generally find that radial gradients are steeper for younger stars. As we discuss below, the origin and evolution of these radial gradients are under active debate.
Works measuring vertical gradients for stars in the MW \citep[for example,][]{Hayden2014, Imig2023} find that vertical gradients are negative, such that stars closer to the disk midplane are more enriched. These works also find that vertical gradients generally decrease in strength with increasing radius. Furthermore, observations measured azimuthal variations for stars in the MW \citep[for example,][]{Anders2023, Hawkins2023, Ratcliffe2023}, although the strength of these variations remains debated.

Accurately measuring these spatial and temporal patterns in elemental abundance, combined with accurately modeling their origin and how they changed across cosmic time, can reveal the formation history of the MW via ``chemical tagging".
Introduced in \citet{FreemanBlandHawthron2002}, chemical tagging aims to use the (near) invariance of stars' elemental abundances to tag observed stellar populations to their birth times and locations.
A key open question for chemical tagging is how homogeneous the ISM was across a galaxy's formation history, which sets the information content of elemental abundances in terms of birth location.
If we can understand and model this, then chemical tagging also offers the possibility of measuring the amount of radial and vertical redistribution that stars today experienced since birth.
If the effects of radial and vertical redistribution are small, on average, then the abundance patterns of stars today nearly directly tell us about the evolution of the (star-forming) ISM across a galaxy's history.

One can think of two regimes for chemical tagging.
``Strong" chemical tagging attempts to tie stars to their birth clusters/clouds \citep[for example,][]{Bovy2016, PriceJones2020, Casamiquela2021}.
To be effective, this would require that star-forming gas clouds be both internally homogeneous and sufficiently unique from each other across the galaxy.
``Weak" chemical tagging attempts to tie stars to their general birth time and broad birth location in the galaxy \citep[for example,][]{Anders2017}. In terms of spatial variations, weak chemical tagging requires simply that stars born in a given patch (of some size) in radius, vertical height, and/or azimuth have the same elemental abundances, and that these vary as a function of radius, vertical height, and/or azimuth.

For chemical tagging to be successful, we must have an effective theoretical model for how the spatial trends of mono-age stellar populations today relate to the trends at their formation (birth), through (semi)analytical models \citep[for example,][]{Minchev2018, Sharda2021}, idealized (non-cosmological) $N$-body simulations \citep[for example,][]{Loebman2016, Khoperskov2022}, and/or cosmological simulations \citep[for example,][]{Gibson2013, VK2018, VK2020, Bellardini2021, Bellardini2022, Lu2022b, Khoperskov2022, Buck2023}.
As we discuss in Section~\ref{sub:Compare_to_other_cos_sim}, among these many works, expectations remain mixed as to the strength and evolution of spatial variations of metallicity at birth, particularly concerning whether radial gradients evolved over cosmic time to be stronger (more negative), weaker (less negative), or experienced little evolution.
Furthermore, while many works explored these metallicity patterns in the ISM or for stars at birth, fewer works analyzing cosmological simulations have compared these patterns at birth to the patterns today, which also enables comparisons with observations of the MW and nearby galaxies.

In this work, we use the Feedback In Realistic Environments (FIRE) cosmological zoom-in simulations to explore these trends across the formation histories of MW-mass galaxies.
FIRE-2 simulations are compelling theoretical tools for this study, because they are fully cosmological, explicitly and self-consistently model the key processes for metal production (stellar nucleosynthesis) and turbulent mixing in the ISM, and previous works have shown that they successfully model key features of MW-like disk galaxies: including thick+thin disk morphology \citep{Ma2017b, Garrison-Kimmel2018, Sanderson2020, Yu2023, Gurvich2023}, dynamics of young stars \citep{McCluskey2023} and old metal-poor stars \citep{Santistevan2021}, atomic gas scale heights \citep{Gensior2023}, stellar bars \citep{Ansar2023}, spiral arms \citep{Orr2023}, and giant molecular clouds \citep{Benincasa2020, Guszejnov2020}.

Our work builds on previous works that analyzed spatial variations of metals in both gas and stars in FIRE simulations.
\citet{Ma2017b} showed that a FIRE simulation, m12i, broadly agrees with the MW in terms of radial and vertical metallicity gradients of stars.
\citet{Ma2017a} explored the redshift and mass dependence of metallicity gradients in the FIRE-1 simulations, showing that strong (negative) gradients only develop in galaxies with rotation-dominated dynamics, $v_{\rm \phi} / \sigma_{\rm v} \gtrsim 1$.
\citet{Bellardini2021} used the same suite of MW-mass FIRE-2 simulations we use here to study spatial variations in gas-phase metallicity. Consistent with \citet{Ma2017b}, they found that negative radial gradients become stronger over time, while azimuthal variations get weaker, as the galaxies evolve to more rotation-dominated dynamics, with a transition in which one dominates over the other in the overall metallicity variations across a galaxy $\approx 8 \Gyr$ ago.
\citet{Bellardini2022} showed that these trends persist for stars at birth, and they also found weak to no mono-age vertical gradients in stars at birth. The transition from azimuthally to radially dominant metallicity variations $\approx 8 \Gyr$ ago coincides with the onset of rotationally-dominated motion ($v_{\rm \phi} / \sigma_{\rm v} \gtrsim 1$) and disk settling that \citet{McCluskey2023} found for these galaxies. \citet{Carillo2023} also showed that this transition time correlates with the beginning of evidence for inside-out radial formation for younger ($< 7$ Gyr) and higher-metallicity ([Fe/H] $> -0.25$ dex) stars. 

In particular, we extend the work of \citet{Bellardini2021} and \citet{Bellardini2022}, to investigate spatial variations of stellar elemental abundances both today and at birth using the FIRE-2 cosmological zoom-in simulations. Importantly, we also compare with various observations of stars in the MW.

\section{Methods}
\label{sec:Methods}

\subsection{FIRE-2}
\label{sec:Metallicity}

\begin{table*}
\centering
\begin{tabular}{>{\centering\arraybackslash\hspace{0pt}} p{22 mm} | >{\centering\arraybackslash\hspace{0pt}} p{15 mm} >{\centering\arraybackslash\hspace{0pt}} p{15 mm} >{\centering\arraybackslash\hspace{0pt}} p{15 mm} >{\centering\arraybackslash\hspace{0pt}} p{20 mm} >{\centering\arraybackslash\hspace{0pt}} p{20 mm} >{\centering\arraybackslash\hspace{0pt}} p{20 mm} >{\centering\arraybackslash\hspace{0pt}} p{25 mm}}
\hline
simulation & $M^{\rm star}_{90}$ & $R^{\rm star}_{90}$ & $Z^{\rm star}_{90}$ & $\partial {\rm [Fe/H]} / \partial R$ & $\sigma_{\rm [Fe/H]}$ & $\partial {\rm [Fe/H]} / \partial |Z|$ & Lookback time \\
 & & & & $(< 1 \Gyr)$ & $(< 1 \Gyr)$ & $(R = 8 \kpc)$ & of disk onset \\
& $[10^{10} {\rm M_\odot}]$ & ${\rm [kpc]}$ & ${\rm [kpc]}$ & [dex kpc$^{-1}$] & [dex] & [dex kpc$^{-1}$] & [Gyr] \\
\hline
m12m$^1$ & 10.0 & 11.9 & 2.3 & $-0.031$ & $0.063$ & $-0.11$ & $9.21$\\
Romulus$^2$ & 8.0 & 14.8 & 2.4 & $-0.034$ & $0.078$ & $-0.067$ & $7.42$\\
m12b$^3$ & 7.3 & 9.2 & 1.8 & $-0.031$ & $0.053$ & $-0.086$ & $7.42$\\
m12f$^4$ & 6.9 & 13.5 & 2.1 & $-0.032$ & $0.084$ & $-0.058$ & $7.42$\\
Thelma$^3$ & 6.3 & 11.7 & 3.2 & $-0.026$ & $0.063$ & $-0.045$ & $4.35$\\
Romeo$^3$ & 5.9 & 14.2 & 1.9 & $-0.041$ & $0.084$ & $-0.14$ & $11.0$\\
m12i$^5$ & 5.3 & 10.1 & 2.3 & $-0.030$ & $0.050$ & $-0.069$ & $6.65$\\
m12c$^3$ & 5.1 & 9.2 & 2.0 & $-0.025$ & $0.058$ & $-0.045$ & $6.49$\\
Remus$^2$ & 4.0 & 12.4 & 2.2 & $-0.038$ & $0.066$ & $-0.14$ & $7.93$\\
Juliet$^3$ & 3.3 & 9.5 & 2.2 & $-0.049$ & $0.12$ & $-0.046$ & $4.35$\\
Louise$^3$ & 2.3 & 12.6 & 2.2 & $-0.047$ & $0.089$ & $-0.12$ & $7.16$\\
\hline
mean & 5.9 & 11.7 & 2.2 & $-0.035$ & $0.074$ & $-0.084$ & $7.22$\\
\hline
\end{tabular}
\caption{
Stellar properties today of the 11 FIRE-2 galaxies in our analysis, in decreasing order of stellar mass. From left to right, the columns are: the name of the simulation; $M^{\rm star}_{90}$ is the stellar mass within $R^{\rm star}_{\rm 90}$; $R^{\rm star}_{\rm 90}$ is the radius that encloses 90\% of stars; $Z^{\rm star}_{\rm 90}$ is the vertical height that encloses 90\% of stars; the radial gradient of \textit{young} stars (ages $< 1 \Gyr$); the azimuthal scatter ($1 \sigma$ standard deviation) of \textit{young} stars (ages $< 1 \Gyr$) at $R = 8 \pm 0.5 \kpc$; the vertical gradient of \textit{all} stars at $R = 8 \pm 0.5 \kpc$; the onset of disk formation, defined in Table~1 of \citet{McCluskey2023} via $v_{\rm \phi} / \sigma_{\rm tot} > 1$ at the time of formation.
The publication that introduced each simulation is: \citet{Hopkins2018}$^1$, \citet{GarrisonKimmel2019b}$^2$, \citet{GarrisonKimmel2019a}$^3$, \citet{GarrisonKimmel2017}$^4$, \citet{Wetzel2016}$^5$.
}
\label{tab:galaxy_properties}
\end{table*}

We analyze 11 Milky Way-mass galaxies from two suites of cosmological zoom-in simulations from FIRE-2 \citep{Hopkins2018}: Latte and ELVIS (Exploring the Local Volume in Simulations).
These FIRE-2 simulations are publicly available \citep{Wetzel2023}.
From the Latte suite, we include 5 isolated galaxies (m12b, m12c, m12f, m12i, m12m) with halo masses of $1 - 2 \times 10^{12}$ M$_\odot$, dark-matter mass resolution of $\approx 3.5 \times 10^{5}$ M$_\odot$, and initial baryonic particle mass of $7070$ M$_\odot$; because of stellar mass loss, most star particles have $\approx 5000$ M$_\odot$.
The ELVIS suite consists of 3 Local Group-like galaxy pairs (Romeo \& Juliet, Romulus \& Remus, and Thelma \& Louise). Their mass resolution is $\approx 2 \times$ better than Latte, with initial baryonic particle masses of $3500 - 4000$ M$_\odot$.
Star and dark matter particles have fixed gravitational softenings, with a Plummer equivalent of $\epsilon_{\rm star} = 4$ pc and $\epsilon_{\rm dm} = 40$ pc, comoving at $z > 9$ and physical at $z < 9$. Gas cells use fully adaptive gravitational softening, which matches the hydrodynamic kernel smoothing, reaching a minimum of 1 pc.

We generated cosmological zoom-in initial conditions using MUSIC \citep{HahnAbel2011}.
Each zoom-in region is embedded within a cosmological box with side lengths of $70.4 - 172$ Mpc. The simulations assume a flat $\Lambda$CDM cosmological model with parameters broadly aligned with \citet{PlanckCollab2020}: $h = 0.68 - 0.71, \Omega_{\Lambda} = 0.69 - 0.734, \Omega_{m} = 0.266 - 0.31, \Omega_{b} = 0.0455 - 0.048, \sigma_{s} = 0.801 - 0.82,$ and $n_{s} = 0.961 - 0.97$.
Each simulation stores 600 snapshots, spanning $z = 99$ to $z = 0$, with time spacing $\lesssim 25$ Myr.

We ran these simulations using the \textsc{Gizmo} code \citep{Hopkins2015}, with the mesh-free finite-mass (MFM) mode for hydrodynamics, a quasi-Lagrangian Godunov method that provides adaptive spatial resolution while maintaining exact conservation of mass, energy, and momentum, excellent angular momentum conservation, and accurate capturing of shocks.
For gravity, \textsc{Gizmo} uses an improved version of the Tree-PM solver from GADGET-2 \citep{Springel2005}.

Our simulations use the FIRE-2 physics model \citep{Hopkins2018}, with the redshift-dependent, spatially-uniform cosmic ultraviolet background from \citet{FaucherGiguere2009}.
FIRE-2 accounts for the metallicity-dependent radiative heating and cooling processes for gas across $10-10^{10}$ K. These processes include free-free, photoionization and recombination, Compton, photo-electric and dust collisional, cosmic ray, molecular, metal-line, and fine-structure processes, accounting for 11 elements (see below).
A gas cell becomes a star particle on a local gravitational free-fall time when it becomes self-gravitating, Jeans-unstable, cold ($T < 10^{4}$ K), and molecular \citep{KrumholzGnedin2011}, inheriting the mass and metallicity of its progenitor gas cell.
A star particle represents a mono-age, mono-abundance stellar population, assuming a \citet{Kroupa2001} initial mass function.

\subsection{Metal enrichment in FIRE-2}
\label{sec:Metal_enrichment_in_FIRE2}

Once stars form, FIRE-2 models their feedback and nucleosynthesis.
Stellar feedback includes winds, core-collapse and white-dwarf (Ia) supernovae, photoionization, photo-electric heating, and radiation pressure. For stellar winds and their yields, FIRE-2 uses \citet{Wiersma2009}, which synthesized a combination of models \citep{VandenHoek1997, Marigo2001, Izzard2004}. Core-collapse supernova rates are from STARBURST99 \citep{Leitherer1999} and yields are from \citet{Nomoto2006}. White-dwarf (Ia) supernova rates are from \citet{Mannucci2006}, with nucleosynthetic yields from \citet{Iwamoto1999}.
FIRE-2 tracks 11 elements: H, He, C, N, O, Ne, Mg, Si, S, Ca, Fe.
We quote all elemental abundances scaled to (proto) Solar values from \citet{Asplund2009}.
Important for our analysis, FIRE-2 also explicitly models sub-grid mixing and diffusion of metals in gas via turbulent eddies \citep{Su2017, Escala2018, Hopkins2018}.
As \citet{Bellardini2021} showed, the details of this diffusion model do not significantly affect radial or vertical gradients in FIRE-2 MW-mass galaxies, but they strongly affect the azimuthal scatter.
As \citet{Bellardini2021} also showed, these FIRE-2 MW-mass galaxies have azimuthal scatter in gas at $z = 0$ similar to that observed in nearby galaxies of similar mass.

\subsection{Milky Way-mass galaxies in FIRE-2}

Table~\ref{tab:galaxy_properties} presents key properties of stars for our 11 MW-mass FIRE-2 galaxies today.
We include only FIRE-2 galaxies with stellar masses within a factor of $\approx 2$ of the MW.
These galaxies experienced a wide range of formation histories \citep{Santistevan2020, McCluskey2023}, given that their selection was agnostic to any properties beyond dark-matter halo mass, including formation history, concentration, spin, or satellite/subhalo population (with the additional constraint of a LG-like paired-halo environment for the ELVIS galaxies).
For key results, we show the mean, $1 \sigma$ scatter, and full range across these 11 MW-mass galaxies.
Therefore, our results should represent trends typical for the formation histories of MW-mass galaxies.

That said, we also compare with observations of the MW, and many observational analyses suggest that the MW's disk began to form early, $\approx 12 \Gyr$ ago \citep[for example,][]{Belokurov2022, Conroy2022, Xiang2022}, which may have influenced its metallicity patterns.
Therefore, throughout we also show key results separately for Romeo, the galaxy within our FIRE-2 sample whose disk began to form the earliest and thus nearest in time to the MW.
This is likely because Romeo formed in a denser, LG-like environment \citep{Garrison-Kimmel2018, Santistevan2020}.
As \citet{McCluskey2023} showed, Romeo's $v_{\rm \phi} / \sigma_{\rm v}$ rapidly increased $11 - 12 \Gyr$ ago, likely within $\approx 1 \Gyr$ of the MW.

\subsection{Selecting and measuring stars}
\label{sec:selection}

When we measure stars today, we include only ``in-situ" stars that formed within $30 \kpc$ comoving of each MW-mass galaxy, and when we measure stars at formation, we include only those that remain within $30 \kpc$ of the galaxy today, following \citet{Bellardini2022}.
These criteria help ensure that we measure similar populations of stars today and at birth, to compare them more fairly.

We include only stars within a vertical range of $|Z| < 3 \kpc$, to constrain our analysis to the disk. We also tested using $|Z| < 1 \kpc$ and $|Z| < 5 \kpc$ and found that this caused negligible ($> 0.001$ dex kpc$^{-1}$) differences in our results.

To calculate the spatial gradient of any profile (radial or vertical), we fit a single linear relation using numpy.polyfit.
However, the radial gradient is sensitive to the radial range. As \citet{Bellardini2022} showed for these FIRE-2 galaxies, the gradient is steeper in the inner galaxy and shallower in the outer galaxy, with an average break radius of $R \approx 5.5 \kpc$. Therefore, quantifying the galaxy via a single gradient is simplified.
However, it allows us to readily examine trends across cosmic time, compare gradients today versus at formation, and average across our galaxy sample.
Throughout, we measure the radial gradient across $R = 0 - 15 \kpc$ today and $R = 0 - R^{\rm star}_{\rm 90}$ at formation (birth).
We tested the results of using different radial ranges: reducing the range today to $R = 0 - 12 \kpc$ steepens the average gradient for young ($> 1$ Gyr) stars from $-0.035$ dex kpc$^{-1}$ to $-0.040$ dex kpc$^{-1}$, while increasing the range to $R = 0 - 20 \kpc$ shallows the average today to $-0.028$ dex kpc$^{-1}$.

In computing any property for a given galaxy, we measure the median across the stellar population at a given time and spatial location on a galaxy-by-galaxy basis.
We then compute the mean, 68\% scatter, and total scatter across our 11 MW-mass galaxies, which is what we show in figures.

We measure all quantities today and at birth using a star particle's instantaneous $R$ and $Z$. Some observational (and theoretical) works, including some MW analyses that we compare against, instead attempt to infer a more general orbital radius, such as the ``guiding center'' radius (under the epicyclic approximation).
In principle, this introduces a source of inconsistency in comparing metallicity profiles.
However, as Bellardini et al. (in prep.) shows for these FIRE-2 galaxies, the results for population statistics that we examine, like radial gradients, do not depend much on the metric of orbital radius, including instantaneous radius.

We focus on [Fe/H], because Fe is the most commonly measured elemental abundance in stars. In Appendix~\ref{sec:appendix1} we also show trends for [C/H] (which are particularly sensitive to stellar winds) and [Mg/H] (which are produced almost entirely in core-collapse supernovae). While spatial gradients in [Mg/H] are slightly weaker than in [Fe/H], and gradients in [C/H] are slightly stronger, gradients for all three are qualitatively similar.
Therefore, our key results are consistent for all elemental abundances in FIRE-2.

\section{Radial profile}
\label{sec:radial_profile}

\subsection{Results in FIRE-2}
\label{sec:Radial_FIRE}

\begin{figure}
\centering
\includegraphics[width = \columnwidth]{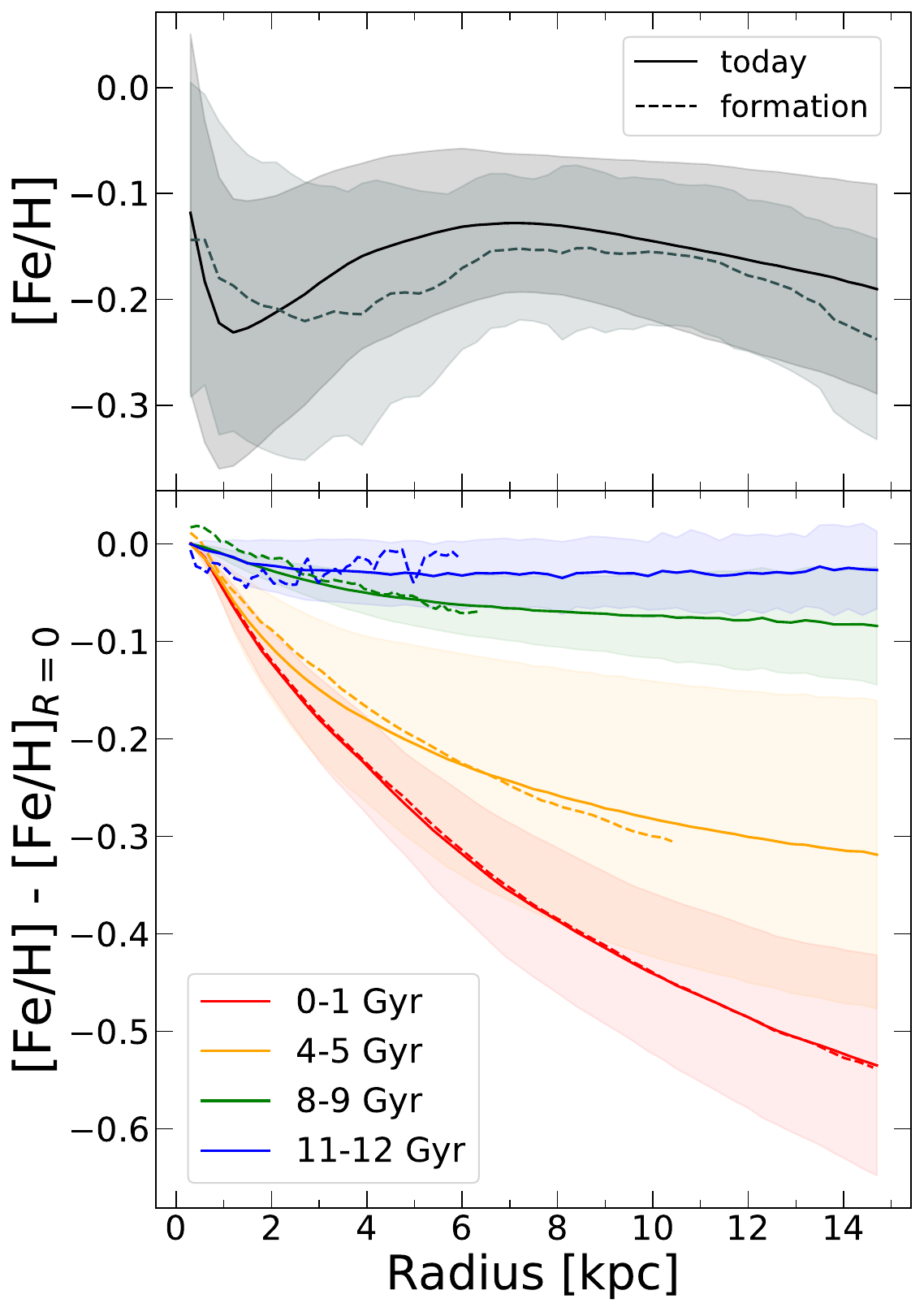}
\caption{\textbf{Top}: Median [Fe/H] of all stars, regardless of age, versus cylindrical radius, $R$. Each line and shaded region show the average and $1 \sigma$ scatter across the 11 galaxies. The solid line shows stars today, while the dashed line shows stars at formation (using their birth radii).
The radial profile for all stars today is nearly flat, though with a dip at $R \approx 2$ kpc from the high density of older stars there. The shape today is similar to that using the birth radii, so radial redistribution of stars does not significantly change this overall radial profile.
\textbf{Bottom}: Median [Fe/H] versus cylindrical radius, $R$, for stars in bins of age, normalized to $\rm [Fe/H] = 0$ at $R = 0$ today. We measure this profile only out to $R^{star}_{90}$ at formation. Radial gradients are always negative at a given age, and they steepened over time, being strongest for stars that formed most recently. Because of radial redistribution, these gradients are marginally shallower today than at formation, but the difference is small.
\textit{Thus, the metallicity profile today and its age dependence primarily reflect the trends already in place when stars formed.}
}
\label{fig:RAD_profile_allstars_AND_RAD_normalized_profile_2plots_PUBLISH}
\end{figure}

Figure~\ref{fig:RAD_profile_allstars_AND_RAD_normalized_profile_2plots_PUBLISH} (top) shows the profile of [Fe/H] for all stars, regardless of age, versus cylindrical radius, $R$.
We show the profile using both the radii of stars today and their radii at birth; the results are qualitatively similar.
[Fe/H] is highest in the galactic center and decreases with $R$ within the inner $2 \kpc$. [Fe/H] then increases between $2 - 6 \kpc$, before decreasing with $R$ beyond that. Observations of the MW also exhibit this increase in [Fe/H] between $2 - 6 \kpc$ and decrease at larger $R$ \citep{Lian2023}. These trends relate to trends in stellar age and the ``inside-out" radial growth histories of these galaxies. As \citet{Bellardini2022} showed (their Figure~1), [Fe/H] increases with decreasing stellar age in FIRE-2.
Furthermore, the inner galaxy contains a wide range of stellar ages (both young and old), while the outer disk contains a much narrower range of stellar ages, primarily just young.
Therefore, we also measure median stellar age as a function of $R$ (not shown): young stars high in [Fe/H] dominate the galactic center, while old stars lower in [Fe/H] dominate the inner regions coinciding with the dip in [Fe/H] in these regions, and then at $R \gtrsim 3 \kpc$ stellar age steadily decreases with $R$.
Given the relatively narrow range of ages in the outer disk, it most directly traces the mono-age radial gradients that we show below. However, trends at $R \lesssim 8 \kpc$ primarily reflect age dependence, and as such, the dip in [Fe/H] at $R \approx 1 - 3 \kpc$ primarily reflects a ``bump" in stellar age.

Because metallicity depends strongly on stellar age, we separate the radial profile into mono-age populations, to uncover the underlying trends across the galaxies' histories.
Figure~\ref{fig:RAD_profile_allstars_AND_RAD_normalized_profile_2plots_PUBLISH} (bottom) shows the radial profile of [Fe/H] for mono-age stars, using age bins of 1 Gyr.
We normalize all profiles to [Fe/H] = 0 at $R = 0 \kpc$ today, to compare the shapes of the profiles.
[Fe/H] declines more significantly with $R$ for younger stars.
This trend is statistically monotonic with age across our sample, and it is similar for stars based on their radii today and at birth.

These radial gradients for stars at birth reflect the conditions of the ISM at that time.
As \citet{Ma2017a} and \citet{Bellardini2021} discussed for FIRE galaxies, these radial gradients emerge from two effects, both of which must be met to cause a strong radial gradient.
First, to \textit{source} a metallicity gradient, smaller radii must have higher star-formation efficiency, that is, mass in young stars (which source most of the new metals) relative to mass in gas (over which the metals are diluted), which drives [Fe/H] to be higher there.
Indeed, as \citet{Bellardini2022} showed for these FIRE-2 galaxies, the radial profile of [Fe/H] for young stars today closely tracks the profile of $\Sigma_{\rm star}(< 1 \Gyr) / \Sigma_{\rm gas}$.
Second, to \textit{sustain} a radial gradient in the ISM, the galaxy must have sufficiently well-ordered rotation (in a disk) to prevent the radial mixing of metals in gas. As these MW-mass galaxies evolved, their ISM developed thinner disks with more ordered rotation \citep{McCluskey2023}, thereby reducing radial mixing, leading to a stronger radial gradient for younger stars.

As Table~\ref{tab:galaxy_properties} shows, for young stars (ages $< 1 \Gyr$), the average gradient across all 11 FIRE-2 galaxies is $-0.035$ dex kpc$^{-1}$, ranging from $-0.026$ dex kpc$^{-1}$ (Thelma) to $-0.049$ dex kpc$^{-1}$ (Juliet).
Noteworthily, the galaxy with the earliest-forming disk, Romeo, has the third steepest radial gradient for young stars of $-0.041$ dex kpc$^{-1}$.
\citet{Bellardini2022} explored the correlation between the radial gradient today and various metrics of formation history for these FIRE-2 galaxies, finding a weak correlation (with significant scatter), such that galaxies that form earlier tend to have stronger (more negative) radial gradients in young stars today.

\citet{Bellardini2022} presented these gradients for stars at birth in these same FIRE-2 galaxies.
Our focus here is to present similar results for stars today and compare the two.
As Figure~\ref{fig:RAD_profile_allstars_AND_RAD_normalized_profile_2plots_PUBLISH} (bottom) shows, the profiles are generally shallower today than at birth, a result of radial redistribution, as Bellardini et al. (in prep.) quantifies.
However, the differences between birth and today are small, with a typical difference of $\approx 0.0053$ dex kpc$^{-1}$, and the largest average difference being $0.013$ dex kpc$^{-1}$ for $6$ Gyr old stars.
\textit{Therefore, the radial profiles of stellar metallicity today primarily reflect those in place in the ISM from which the stars formed.}

\begin{figure}
\centering
\includegraphics[width = \columnwidth]{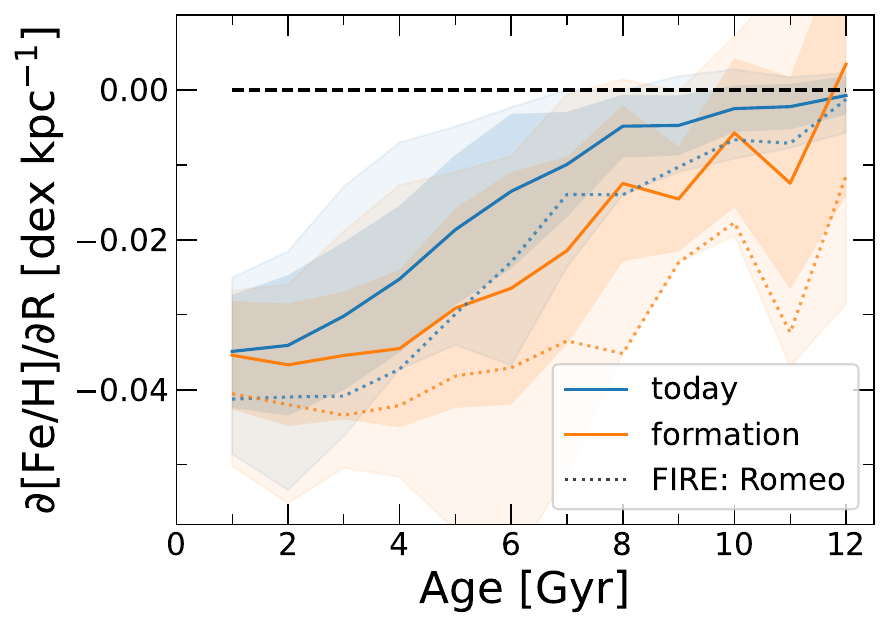}
\caption{Radial gradient of [Fe/H], $\partial {\rm [Fe/H]} / \partial R$, versus stellar age, for stars today and at formation (using their birth radii).
For each galaxy, we measure the gradient across $R = 0 - 15$ kpc today and across $R = 0 - R^{star}_{90}$ at formation.
Each line, darker shaded region, and lighter shaded region show the average, $1 \sigma$ scatter, and min/max across the 11 galaxies. Dotted lines show Romeo, the earliest-forming disk in our sample. 
The radial gradient is almost always negative, both at formation and today, becoming steeper for stars that formed more recently, and being nearly flat for stars that formed $\gtrsim 10 \Gyr$ ago.
Because of radial redistribution after birth, the gradient is marginally shallower today than at formation, but the difference is modest, typically $\lesssim 0.01$ dex kpc$^{-1}$, and the trends with age are unchanged.
\textit{Thus, the radial gradient of metallicity today primarily reflects the conditions of stars at birth.}
}
\label{fig:RAD_slope_PUBLISH}
\end{figure}

Figure~\ref{fig:RAD_slope_PUBLISH} quantifies these radial gradients by showing $\partial {\rm [Fe/H]} / \partial R$ as a function of stellar age.
Again, we measure a linear fit across $R = 0 - 15 \kpc$ today and $R = 0 - R^{\rm star}_{\rm 90}$ at formation. 
As in Figure~\ref{fig:RAD_profile_allstars_AND_RAD_normalized_profile_2plots_PUBLISH}, this gradient flattens for older stars, both when using radii today and radii at birth.
At ages $\lesssim 3 \Gyr$, the radial gradient at birth was constant with age, as \citet{Bellardini2022} showed.
As expected, radial redistribution causes the radial gradient today to be systematically shallower (smaller) than at formation.
Interestingly, this difference increases with age up to $\approx 3 \Gyr$, but the difference between birth and today, $\approx 0.01$ dex kpc$^{-1}$, is nearly constant with age for stars older than this.
This is consistent with Bellardini et al. (in prep.) that the amount of radial redistribution in these galaxies increases with age up to $\approx 3 - 4 \Gyr$ but no longer depends much on age beyond that. 

In Figure~\ref{fig:RAD_slope_PUBLISH}, we also show Romeo, which has systematically steeper radial gradients than the mean trend at all ages, as a result of its early-forming disk, which we further discuss in the next section, and its pertinence to radial gradients in the MW.

\subsection{Comparison to the Milky Way}
\label{sec:radial_comparison}

\citet{Bellardini2021} found that the typical radial gradient of metallicity in gas in these FIRE-2 galaxies today is $\partial {\rm [O/H]} / \partial R = -0.028$ dex kpc$^{-1}$.
This agrees well with gradients measured for M31, from \citet{Zurita2012} and \citet{Sanders2012} ($-0.023$ dex kpc$^{-1}$ and $-0.0195$ dex kpc$^{-1}$, respectively), and is steeper than observed in most nearby MW-mass galaxies, but this is shallower than most measurements of the MW of $\approx -0.046$ dex kpc$^{-1}$ \citep[see references in][]{Bellardini2021}. \citet{Bellardini2022} found similar results for young stars at birth as well.
This suggests that the MW has a steeper radial gradient than many nearby galaxies of similar mass. We discuss this further in Section~\ref{sub:stength_of_rad_grad}.

\begin{figure}
\centering
\includegraphics[width = \columnwidth]{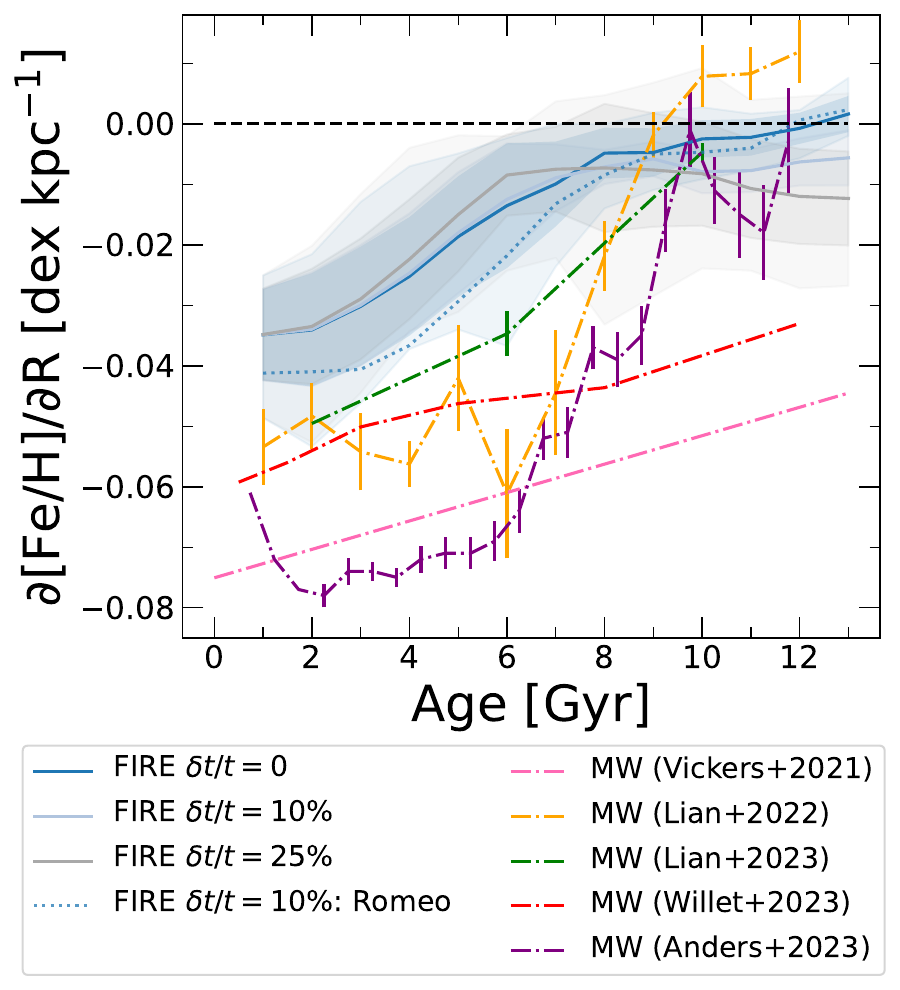}
\caption{Radial gradient of [Fe/H] today, $\partial {\rm [Fe/H]} / \partial R$, versus stellar age, assuming various fractional uncertainties in stellar age, $\delta t / t$. Each line, darker shaded region, and lighter shaded region show the average, $1 \sigma$ scatter, and min/max across the 11 galaxies. The dotted line shows Romeo, the earliest-forming disk in our sample, with $\delta t/t = 10\%$.
The intrinsic radial gradient is monotonically flatter for older stars.
However, the trend with age can reverse to become steeper for older stars for sufficiently large age uncertainties, which mixes younger stars with steeper gradients into older age bins.
We compare with observations of the MW from \citet{Vickers2021}, \citet{Lian2022}, \citet{Lian2023}, \citet{Willet2023}, and \citet{Anders2023}.
FIRE-2 simulations have weaker radial gradients than the MW at most ages, though the steepest gradients in FIRE-2 are nearly consistent with \citet{Lian2023}.
More importantly, the trend with age in FIRE-2 agrees with the MW, and FIRE-2 agrees with \citet{Lian2022}, \citet{Lian2023}, and \citet{Anders2023} that the radial profile becomes nearly flat for the oldest stars.
\textit{In conjunction with Figure} \ref{fig:RAD_slope_PUBLISH}\textit{, these trends imply that the age dependence in the MW primarily reflects the radial gradients of stars at birth.}
}
\label{fig:RAD_slopes_MW_compare_INCOMPLETE}
\end{figure}

Several observational works have measured how the radial gradient of stellar metallicity in the MW depends on stellar age.
Figure~\ref{fig:RAD_slopes_MW_compare_INCOMPLETE} shows $\partial {\rm [Fe/H]} / \partial R$ versus stellar age, 
for five recent observational analyses of the MW, as we summarize below.

\citet{Vickers2021} used LAMOST and Gaia to calculate the ages and orbital radii of $\approx 4$ million stars to determine the radial gradients in [Fe/H] for a sample of 1.3 million stars. Figure~13 in \citet{Vickers2021} shows the radial gradient as a function of age, which we show in Figure~\ref{fig:RAD_slopes_MW_compare_INCOMPLETE}. \citet{Vickers2021} found a radial gradient for the youngest stars of $\partial {\rm [Fe/H]} / \partial R \approx -0.075$ dex kpc$^{-1}$. This gradient decreases with stellar age but remains fairly steep for stars 13 Gyr in age, at $\approx -0.045$ dex kpc$^{-1}$.

\citet{Lian2022} used APOGEE DR16 to study mono-age populations of stars. Using lines of best fit, we convert the radial profiles in Figure~A2 of \citet{Lian2022} which span $0 \kpc \lesssim R \lesssim 13 \kpc$ into radial gradients as a function of age, and we show the results in Figure~\ref{fig:RAD_slopes_MW_compare_INCOMPLETE}. \citet{Lian2022} found a radial gradient for stars $< 1 \Gyr$ of $\partial {\rm [Fe/H]} / \partial R \approx -0.053$ dex kpc$^{-1}$. Their gradient becomes shallower with stellar age and even becomes positive for stars older than 9 Gyr.

\citet{Lian2023} used APOGEE and Gaia to determine metallicity profiles of mono-age populations. Here, they divided stars into larger bins of age: $0 - 4 \Gyr$, $4 - 8 \Gyr$, and $8 - 12 \Gyr$ and measured profiles over a range of $2 \kpc \lesssim R \lesssim 14 \kpc$. We show the gradients of the profiles from Figure~1 of \citet{Lian2023} in Figure~\ref{fig:RAD_slopes_MW_compare_INCOMPLETE}. \citet{Lian2023} found a radial gradient for stars aged $0 - 4 \Gyr$ of $\partial {\rm [Fe/H]} / \partial R \approx -0.050$ dex kpc$^{-1}$, which flattens to $\approx -0.035$ dex kpc$^{-1}$ for stars $4 - 8 \Gyr$ old and further flattens to $\approx -0.005$ dex kpc$^{-1}$ for stars $8 - 12 \Gyr$ old.

\citet{Willet2023} used APOGEE DR17 and Gaia DR3 to investigate how metallicity changes with radius in the thin disk using a sample of 668 red-giant stars. They divided stars into age bins of $0-1$ Gyr, $1-2$ Gyr, $2-4$ Gyr, $4-6$ Gyr, $6-10$ Gyr, and $> 10 \Gyr$, and they presented [Fe/H] profiles across $4 \kpc < R < 11 \kpc$ in their Figure~4, of which we show the gradients in Figure~\ref{fig:RAD_slopes_MW_compare_INCOMPLETE}. \citet{Willet2023} found a radial gradient for stars aged $0 - 1 \Gyr$ of $\partial {\rm [Fe/H]} / \partial R \approx -0.059$ dex kpc$^{-1}$. This gradient flattens with stellar age but still remains significant for stars $> 10 \Gyr$ at $\approx -0.033$ dex kpc$^{-1}$.

\citet{Anders2023} used APOGEE to estimate stellar ages and investigate elemental, positional, and kinematic relationships of stars. Figure~11 of \citet{Anders2023} shows the radial gradient of [Fe/H], measured using Galactocentric distance across $5 \kpc < R < 11 \kpc$, versus stellar age, which we show in Figure~\ref{fig:RAD_slopes_MW_compare_INCOMPLETE}. \citet{Anders2023} found a radial gradient of $\partial {\rm [Fe/H]} / \partial R \approx -0.061$ dex kpc$^{-1}$ for stars aged $> 1 \Gyr$, which increases to $\approx -0.078$ dex kpc$^{-1}$ at ages $\approx 2.25$ Gyr. For ages $\gtrsim 2.25 \Gyr$, this gradient generally decreases with stellar age, becoming essentially flat for stars aged $\approx 11.75$ Gyr.

Figure~\ref{fig:RAD_slopes_MW_compare_INCOMPLETE} compares these observational analyses.
They all agree that the radial gradient becomes shallower for older stars in the MW.
However, they show a significant amount of variation at a given age.
For stars younger than $\approx 2 \Gyr$, the measured gradient varies from $\approx -0.05$ dex kpc$^{-1}$ to $\approx -0.075$ dex kpc$^{-1}$.
These works find even wider variations for the oldest stars.
\citet{Lian2022}, \citet{Lian2023}, and \citet{Anders2023} find that the gradient becomes essentially flat for stars older than $\approx 10 \Gyr$, while \citet{Vickers2021} and \citet{Willet2023} find much weaker age dependence, such that the gradient remains steep at these older ages.
These works use different surveys, with different selection functions, and different approaches to measuring stellar ages, all of which likely contribute to this scatter.

Figure~\ref{fig:RAD_slopes_MW_compare_INCOMPLETE} also shows $\partial {\rm [Fe/H]} / \partial R$ today versus stellar age from our FIRE-2 galaxies.
Given that observational measurements of the ages of stars generally have non-trivial uncertainties, we also assume various fractional uncertainties in stellar age ($\delta t / t$) in measuring this for our FIRE-2 galaxies, aiming to bracket age uncertainties typical of these observational surveys, $\gtrsim 10 - 30 \%$ \citep[for example,][]{Soderblom2010, SilvaAguirre2018, Shen2024}. Furthermore, Figure~\ref{fig:RAD_slopes_MW_compare_INCOMPLETE} shows (via a dotted line) the trend for Romeo, our earliest-forming disk and likely the closest in time to when the MW disk spun up, which shows steeper than average gradients at nearly all ages.
Compared with measurements of the MW, the radial gradients in FIRE-2 galaxies are generally weaker at a given age. We discuss potential reasons for this in Section~\ref{sub:stength_of_rad_grad}. The steepest gradient in FIRE-2 ($\partial {\rm [Fe/H]} / \partial R = -0.049$ dex kpc$^{-1}$ for young stars) is consistent with \citet{Lian2023} across most of the age range, being $0.050$ dex kpc$^{-1}$ for their youngest age bin.
However, all other analyses of the MW in Figure~\ref{fig:RAD_slopes_MW_compare_INCOMPLETE} are steeper than all of our FIRE-2 galaxies. That said, FIRE-2 agrees with \citet{Lian2022}, \citet{Lian2023}, and \citet{Anders2023} that the radial gradients are essentially flat for the oldest stars, and, most importantly, the general trend of the radial gradient becoming less steep (flatter) with increasing stellar age agrees with \textit{all} of these observations.

Figure~\ref{fig:RAD_slopes_MW_compare_INCOMPLETE} also highlights the effects of applying uncertainties in stellar age to FIRE-2 galaxies.
As expected, larger uncertainties mix stars of different intrinsic ages into the same ``measured" age bin, thus mixing stars that formed with different intrinsic radial gradients.
The effect of this mixing depends on the slope of the underlying relation between the gradient and age.
For stars younger than $\approx 8 \Gyr$ in FIRE-2, larger age uncertainties act to shallow the gradient at a given ``measured" age, while for older stars, which formed with little-to-no intrinsic gradients, this mixing pulls in younger stars and leads to a steeper ``measured" gradient.
Therefore, stellar age uncertainties may contribute to the widely varying gradients observed for old stars in the MW.
As Figure~\ref{fig:RAD_slopes_MW_compare_INCOMPLETE} shows, fractional age uncertainties of 10\% do not affect this relation much, but uncertainties of 25\% or larger bias the relation significantly.

Finally, because the radial trends in FIRE-2 today are broadly similar to those in the MW in Figure \ref{fig:RAD_slopes_MW_compare_INCOMPLETE}, and these trends in FIRE-2 were mostly set at birth, this suggests that the radial trends of metallicity in the MW today were set mostly at birth as well.

\section{Vertical profile}
\label{sec:Vertical_Profile}

\subsection{Results in FIRE-2}
\label{sec:Vertical_FIRE}

\begin{figure}
\centering
\includegraphics[width = \columnwidth]{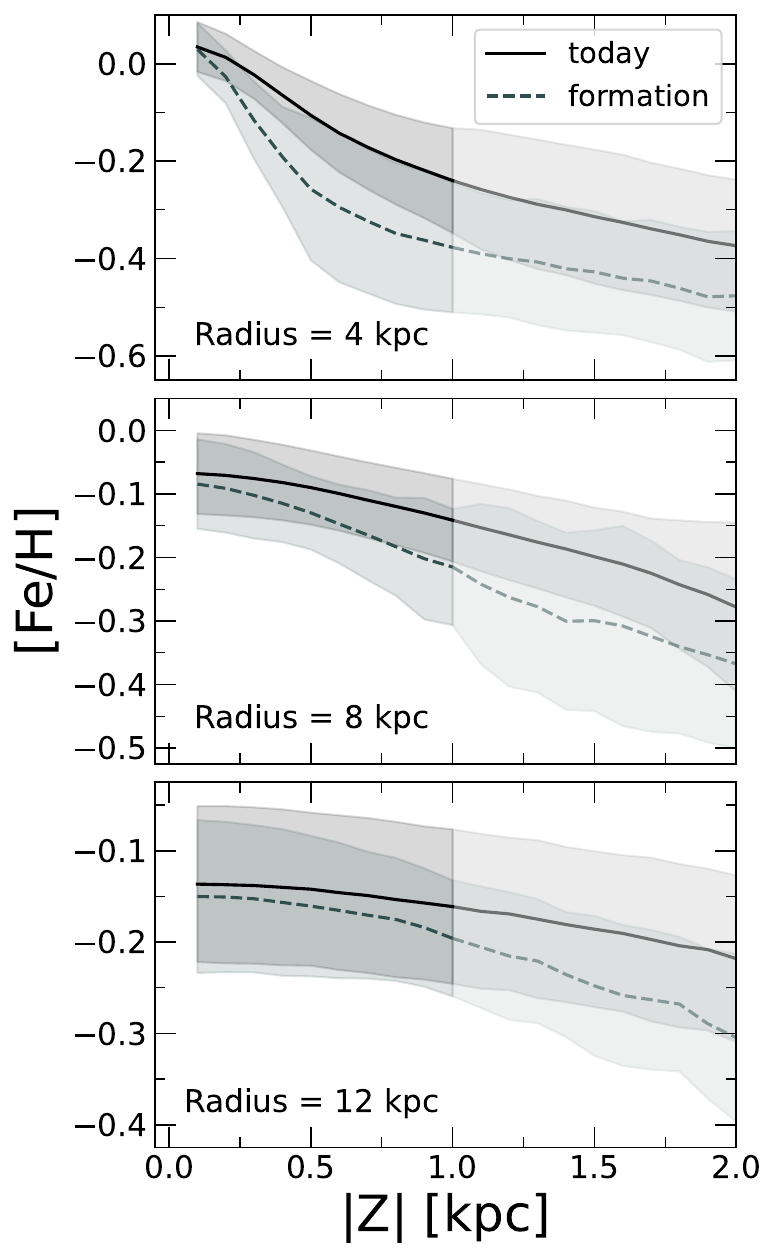}
\caption{[Fe/H] for all stars versus absolute vertical height above/below the disk midplane, $|Z|$, both today (using $|Z|$ and $R$ today) and at formation (using $|Z|$ and $R$ at birth), at different $R$ ($\pm 0.5 \kpc$): 4 kpc (inner disk), 8 kpc (Solar annulus), and 12 kpc (outer disk).
Each line and shaded region show the average and $1 \sigma$ scatter across the 11 galaxies.
[Fe/H] decreases with increasing vertical height at all $R$, but this decrease is weaker at larger $R$.
\textit{All of these trends are present both today and at formation}: they arise from the ``upside-down" nature of disk settling, and because stars span a larger range of metallicity (and age) at smaller $R$.
As Appendix~\ref{sec:appendix2} shows, the vertical gradients for mono-age populations are weak.
} 
\label{fig:VER_profile_allstars_PUBLISH}
\end{figure}

\begin{figure*}
\centering
\includegraphics[scale = 0.52]{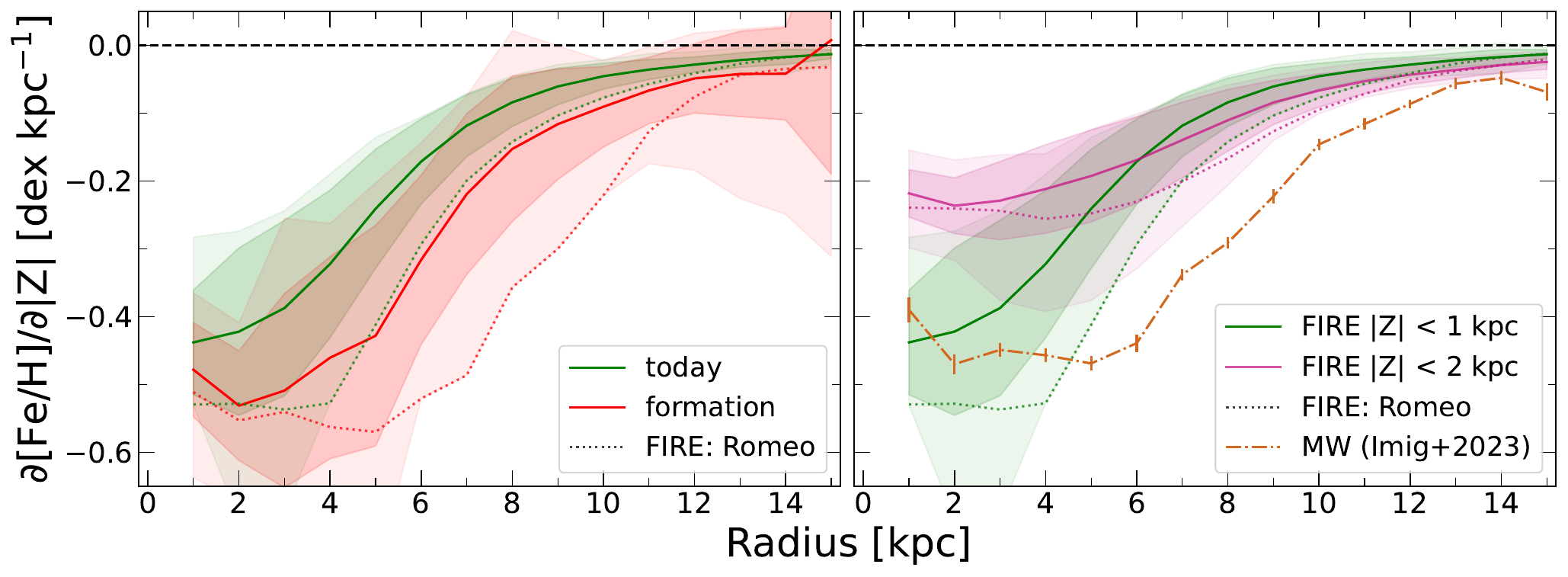}
\caption{Vertical gradient in [Fe/H], $\partial {\rm [Fe/H]} / \partial |Z|$, of all stars regardless of age versus cylindrical radius, $R$. Each line, darker shaded region, and lighter shaded region show the average, $1 \sigma$ scatter, and min/max across the 11 galaxies.
Dotted lines show Romeo, the earliest-forming disk in our sample.
\textbf{Left}: Measuring the vertical gradient across $|Z| = 0 - 1 \kpc$, both today and at formation (using birth $|Z|$ and $R$). The vertical gradient is always negative, both at formation and today, and it steepens at smaller $R$.
As Appendix~\ref{sec:appendix2} shows, the vertical gradients for mono-age populations are weak, so these vertical gradients result from ``upside-down" disk settling, and they are stronger at smaller $R$, where stars span a larger range in metallicity (and age).
The vertical gradients have weakened somewhat between formation and today, from vertical heating and radial redistribution, but these key trends were already in place at birth.
\textbf{Right}: Same, but only for stars today, across a vertical range of $|Z| < 1$ kpc and $|Z|< 2$ kpc, to compare with observations of the MW from \citet{Imig2023}. FIRE-2 has weaker vertical gradients than the MW at most $R$, but the trend of a steepening vertical gradient at smaller $R$ agrees.
This implies that the MW underwent more metal enrichment \textit{as its disk settled} than in FIRE-2 galaxies, likely because FIRE-2 galaxies settled later, when [Fe/H] increased more gradually, than in the MW. 
} 
\label{fig:VER_slope_vs_R_PUBLISH}
\end{figure*}

Figure~\ref{fig:VER_profile_allstars_PUBLISH} shows vertical profiles of [Fe/H] for all stars today, regardless of age, at 3 radii ($\pm 0.5 \kpc$): 4 kpc (inner disk), 8 kpc (Solar annulus), and 12 kpc (outer disk).
We show trends both today and using the positions (both $R$ and $Z$) of stars at their birth. Because some of our galaxies have $|Z|^{\rm star}_{\rm 90} < 2 \kpc$, we lighten the vertical profile at $|Z| > 1 \kpc$, to emphasize that $Z < 1 \kpc$ is our principle range of interest.

[Fe/H] is higher near the midplane of the disk and decreases with absolute vertical height. This trend is strong in the inner disk, weaker at the Solar annulus, and even weaker in the outer disk, but it is consistent throughout. As Appendix~\ref{sec:appendix2} shows, the vertical gradients for mono-age stars are essentially flat, both today and at formation.
Therefore, the primary driver of this vertical profile for all stars today is metal enrichment over time, combined with disk settling over time. The youngest and most metal-rich stars dominate the midplane, while older more metal-poor stars formed (and currently tend to be at) larger vertical height. This results in a vertical gradient in both metallicity and age from ``upside down" disk formation \citep[for example,][]{Bird2013}.

Figure~\ref{fig:VER_slope_vs_R_PUBLISH} (left) quantifies the vertical gradient, $\partial {\rm [Fe/H]} / \partial |Z|$, of all stars regardless of age across $|Z| < 1 \kpc$, versus cylindrical radius, $R$, both today and at formation (using birth $Z$ and $R$). Again, the vertical gradient is strongest at smaller $R$, reaching $\partial {\rm [Fe/H]} / \partial |Z| = -0.44$ dex kpc$^{-1}$ at $R < 1 \kpc$ today and $-0.48$ dex kpc$^{-1}$ based on birth coordinates. At $R = 8 \kpc$, the vertical gradient is $-0.084$ dex kpc$^{-1}$ today and $-0.15$ dex kpc$^{-1}$ using birth coordinates. This gradient weakens with $R$, becoming negligible at $R \gtrsim 10 \kpc$. We also show Romeo, which has systematically steeper vertical gradients than the mean trend at all $R$, which we discuss further in Section \ref{sec:Vertical_Comparison_to_the_Milky_Way}.

This radial dependence for vertical metallicity gradients arises because larger radii systematically have a narrower range of ages (primarily just younger stars), and therefore a narrower range of metallicities (mostly just metal-rich stars), while smaller radii have a wider range of ages that sampled the galaxy across a larger fraction of the disk settling.

Vertical gradients are systematically shallower today than at birth, because the combination of vertical heating and radial redistribution tends to wash out (smooth) the gradients at all $R$. Nonetheless, as with radial gradients, the redistribution of stars between formation and today does not systematically change the overall trends in the vertical gradient, with differences between birth and today typically being $\approx 0.073$ dex kpc$^{-1}$, and the largest average difference being $0.19$ dex kpc$^{-1}$ for stars at $R = 5 \kpc$.

As with the radial profiles, we also split these vertical profiles into mono-age populations to analyze their contributions to these gradients.
As Appendix~\ref{sec:appendix2} shows, unlike the radial gradients, which can be strongest for mono-age populations, the vertical gradients of mono-age stars are mostly negligible, being $\lesssim -0.083$ dex kpc$^{-1}$ today and $\lesssim -0.051$ dex kpc$^{-1}$ using formation coordinates.
This relative flatness of the vertical gradient at a given age arises because the height of the star-forming gas disk is largely regulated by turbulence, which also efficiently mixes metals. Thus, star-forming gas is unable to stratify into a metallicity gradient the way it does radially, causing stars of a given age to possess similar metallicities across the vertical extent.

The origin and age dependence of the vertical gradient is therefore nearly opposite that of the radial gradient.
The radial gradient can be strong for mono-age stars but is nearly flat when combining all stars (Figure~\ref{fig:RAD_profile_allstars_AND_RAD_normalized_profile_2plots_PUBLISH}), whereas the vertical gradient is strong when combining all stars but is negligible for mono-age stars.

\subsection{Comparison to the Milky Way}
\label{sec:Vertical_Comparison_to_the_Milky_Way}

Figure~\ref{fig:VER_slope_vs_R_PUBLISH} (right) compares vertical gradients in these FIRE-2 galaxies against measurements of the MW from \citet{Imig2023}, which used APOGEE DR17 to measure
%radial\footnote{Because \citet{Imig2023} did not analyze radial gradients versus age, we did not directly compare to this work for mono-age radial gradients in Figure~\ref{fig:RAD_slopes_MW_compare_INCOMPLETE}.} and 
vertical gradients in metallicity using 66,496 red-giant stars. We include the results from their Figure~16, which shows $\partial {\rm Fe/H]} / \partial |Z|$ for all stars regardless of age versus $R$. \citet{Imig2023} measured the gradient across different vertical ranges at different $R$. At most radii, they used $|Z| \lesssim 1$ kpc, but at $R \approx 7 - 10 \kpc$, they used $Z \lesssim 2 \kpc$.
To bracket these ranges, Figure~\ref{fig:VER_slope_vs_R_PUBLISH} (right) simply shows vertical gradients in FIRE-2 today across both $|Z| < 1 \kpc$ and $|Z| < 2 \kpc$.
As the profiles in Figure~\ref{fig:VER_profile_allstars_PUBLISH} show, using a larger vertical range can lead to a steeper or shallower vertical gradient, depending on $R$.
In FIRE-2, using a larger vertical range leads to a shallower vertical gradient at $R \lesssim 6 \kpc$, but it leads to a slightly steeper gradient at larger $R$.

\citet{Imig2023} found that the steepest vertical gradient in the MW of $\partial {\rm Fe/H]} / \partial |Z| \approx -0.47$ dex kpc$^{-1}$ occurs at $R = 2 - 5 \kpc$, and it decreases at $R \gtrsim 5 \kpc$ to a minimum of $\approx -0.07$ dex kpc$^{-1}$ at $R = 15 \kpc$. Beyond this, the vertical gradient is negligible. At $R \lesssim 3$ kpc, \citet{Imig2023} measured this across $|Z| \lesssim 1 \kpc$, and the vertical gradients in FIRE-2 at $R \lesssim 4 \kpc$ are broadly consistent with \citet{Imig2023} across this range. At larger $R$, however, the vertical gradients in FIRE-2, regardless of the vertical range, are consistently shallower than the MW from \citet{Imig2023}.

Given our results in Appendix~\ref{sec:appendix2}, that the vertical gradients of mono-age populations contribute negligibly to this trend, this discrepancy with the MW most likely reflects differences in metal enrichment history as it relates to disk settling history.
Specifically, the most rapid rate of metal enrichment in these FIRE-2 galaxies occurs early in their formation history (see Figure~1 of \citealt{Bellardini2022}).
If this enrichment history broadly reflects the MW's, and if the MW's disk started to settle earlier than these FIRE-2 galaxies, then the MW would have experienced more vertical settling during this rapid enrichment phase, leading to the MW having a more vertically stratified metallicity distribution. Indeed, as Figure~\ref{fig:VER_slope_vs_R_PUBLISH} (right) shows, our galaxy with the earliest settling disk, Romeo, has a stronger gradient than our mean trend line, more similar to the MW.
From Figure~\ref{fig:VER_slope_vs_R_PUBLISH} (left), another possibility is that the MW experienced less vertical heating and/or radial redistribution than in FIRE-2 (as we explore directly in Bellardini et al. in prep.), such that its vertical gradient today more directly reflects its trends at the time that stars formed.

Despite this discrepancy in the exact strength of the vertical gradient today, the general trend of a steepening vertical gradient with decreasing radius agrees with the MW.
%Again, we speculate that this is a further consequence of the MW's disk starting to form/settle earlier than any of these FIRE-2 galaxies.
Because the vertical trends in FIRE-2 today are broadly similar to those in the MW, and these trends in FIRE-2 were set mostly at birth, this suggests that the vertical trends of metallicity in the MW today were set mostly at birth as well.

\section{Azimuthal scatter}
\label{sec:Azimuthal_Scatter}

\subsection{Results in FIRE-2}
\label{sec:Azimuthal_FIRE}

We measure the azimuthal scatter of [Fe/H], $\sigma_{\rm [Fe/H]}$, as the standard deviation of [Fe/H] around an annulus (360 degrees) at a fixed $R$, with width $\Delta R = \pm 0.5 \kpc$.
To measure azimuthal scatter for stars at birth, we follow \citet{Bellardini2022}: we split each age bin of $1 \Gyr$ into $20$ age bins, each spanning $50 \Myr$, and we compute the azimuthal scatter within each one; then we average across these 20 bins to compute the typical azimuthal scatter within for a given $1 \Gyr$ age bin.
\citet{Bellardini2022} (their Figure~9) examined how azimuthal scatter varies with azimuthal bin size (arclength) for stars at birth.
They found a weak dependence on the arclength, such that the scatter across a local patch (azimuthal length $\lesssim 1 \kpc$) is only marginally smaller ($\lesssim 0.01$ dex) than the scatter across an entire 360-degree azimuth.
Therefore, we measure the total azimuthal scatter across 360 degrees for simplicity.

\begin{figure}
\centering
\includegraphics[width = \columnwidth]{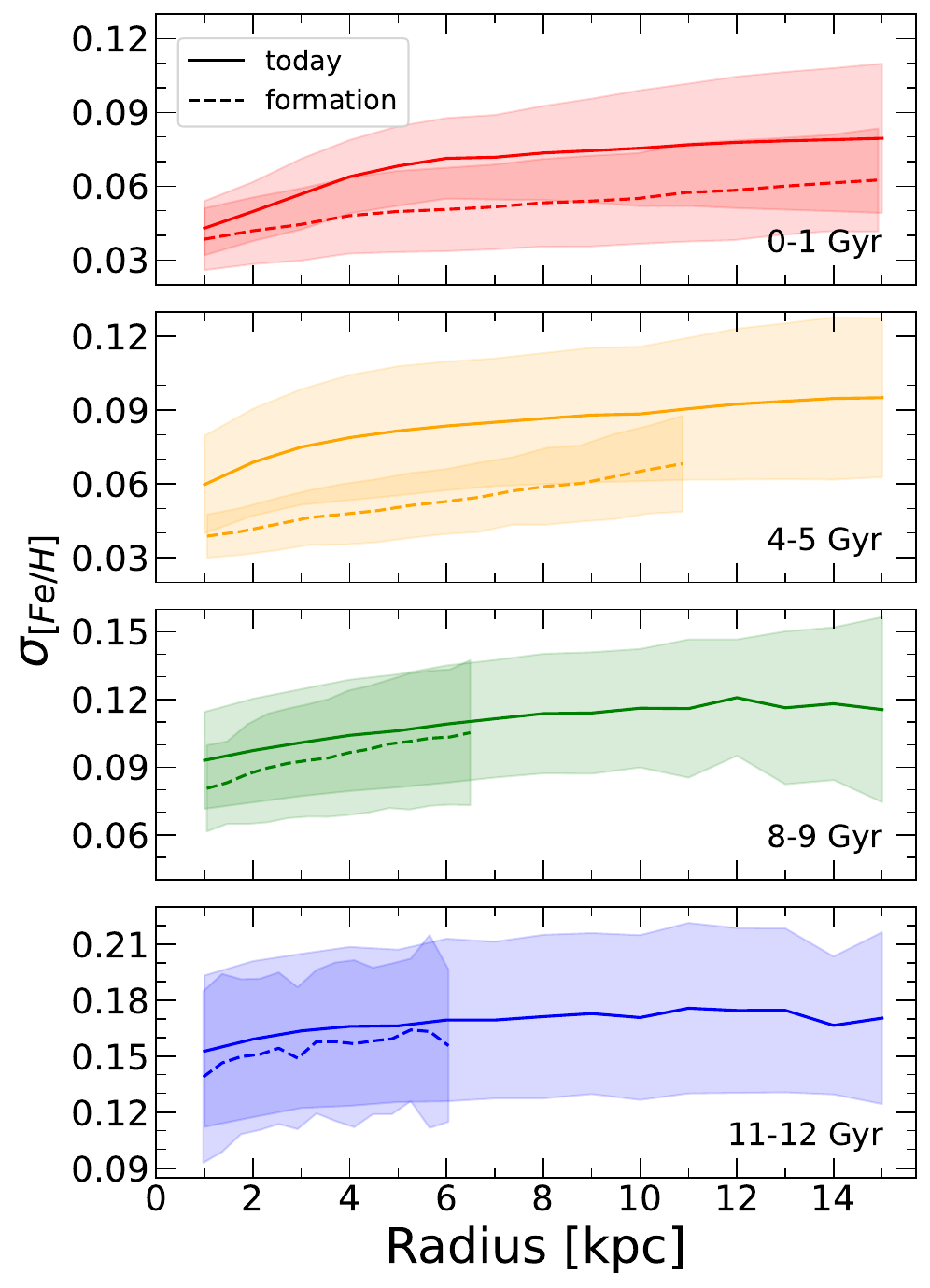}
\caption{Azimuthal scatter in [Fe/H] across an annulus at fixed $R$, $\sigma_{\rm [Fe/H]}$, versus cylindrical radius, $R$, for stars today and at formation (using their birth radii), in bins of stellar age. Each line and shaded region show the average and $1 \sigma$ scatter across the 11 galaxies. We show azimuthal scatter only out to $R^{star}_{90}$ at the time of formation. Azimuthal scatter increases weakly with $R$ at all ages, more so at formation than today.
As a result of radial redistribution, azimuthal scatter is somewhat larger, with weaker radial dependence, today than at birth, but these differences are smaller than $\sigma_{\rm [Fe/H]}$ in place at birth.
\textit{Therefore, most of the azimuthal scatter in stars today was present at their birth.}
} 
\label{fig:AZIM_profiles_4plots_PUBLISH}
\end{figure}

Figure~\ref{fig:AZIM_profiles_4plots_PUBLISH} shows $\sigma_{\rm [Fe/H]}$ versus cylindrical radius, $R$, both today and at formation, in bins of stellar age.
At a given age, $\sigma_{\rm [Fe/H]}$ increases weakly with $R$, in agreement with \citet{Bellardini2022} for stars at birth.
This is because the outer gas disk is less well mixed, from a combination of the longer orbital (and hence mixing) timescales and the stronger influence of cosmic accretion.
This radial dependence is slightly weaker for stars today, as expected from post-formation radial redistribution weakening any radial trends.
For old stars, radial redistribution does not change $\sigma_{\rm [Fe/H]}$ much from formation to today, because there was no strong radial gradient at that time to add to this scatter.
However, for younger stars, radial redistribution acts to increase $\sigma_{\rm [Fe/H]}$ from birth to today, because it adds additional scatter sourced from the radial gradient, as we explore in Section~\ref{sec:relation}.
However, this increase in $\sigma_{\rm [Fe/H]}$ at a given $R$ from birth to today is generally smaller than the $\sigma_{\rm [Fe/H]}$ already in place at birth.
Thus, most of the azimuthal scatter was already in place at the time of stellar birth.

\begin{figure}
\centering
\includegraphics[width = \columnwidth]{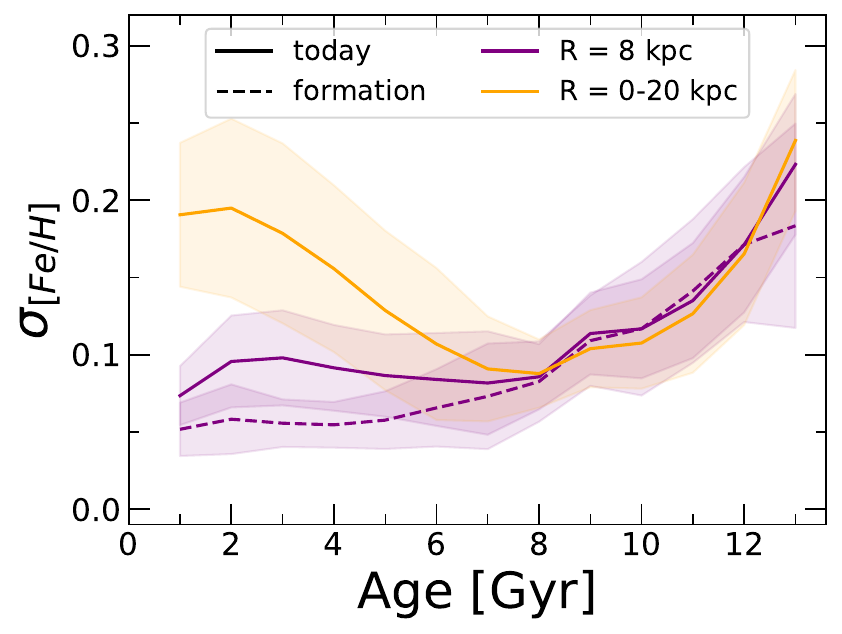}
\caption{Scatter in [Fe/H], $\sigma_{\rm [Fe/H]}$, versus stellar age, for stars today and at formation (using their birth radii). Each line and shaded region show the average and $1 \sigma$ scatter across the 11 galaxies. The purple lines show the azimuthal scatter across the Solar annulus, $R = 8 \pm 0.5 \kpc$, which is nearly constant with age over the last $\approx 8 \Gyr$ and rises rapidly with age before that.
The azimuthal scatter today is only moderately larger than at formation, from the effects of radial redistribution, and only at ages $\lesssim 8 \Gyr$.
\textit{Most of the azimuthal scatter in stars today was present already at their birth.}
For comparison, the gold line shows the \textit{total} scatter in [Fe/H] across the \textit{entire} galaxy, $R < 20 \kpc$, which shows the same trend at old ages, but rises towards younger ages, from the onset of a radial gradient.
}
\label{fig:AZIM_Scatter_vs_Age_PUBLISH}
\end{figure}

Figure~\ref{fig:AZIM_Scatter_vs_Age_PUBLISH} shows $\sigma_{\rm [Fe/H]}$ versus stellar age. The purple lines show the scatter across the Solar annulus ($R = 8 \pm 0.5 \kpc$).
There, the azimuthal scatter was fairly constant for ages $\lesssim 8 \Gyr$, at $\sigma_{\rm[Fe/H]} \approx 0.087$ dex today and $\approx 0.062$ dex at formation.
For stars older than $8 \Gyr$, the azimuthal scatter increases rapidly with age, from increased patchiness of the gas at those times \citep{Bellardini2021, Bellardini2022}. Indeed, \citet{McCluskey2023} found that the disk began forming in these FIRE-2 galaxies $\approx 8 \Gyr$ ago, on average.
Before that time, azimuthal variations via patchiness were the primary cause of metallicity variations across the galaxy.
After that, the radial gradient dominated, as the azimuthal variations smoothed away.
As explained above, for this reason, the azimuthal scatter today versus at birth is nearly identical for stars formed in the ``pre-disk'' era $\gtrsim 8 \Gyr$ ago, while radial redistribution causes the azimuthal scatter today to be somewhat larger than at birth during the ``disk'' era $\lesssim 8 \Gyr$ ago.
That said, as with radial and vertical profiles, the azimuthal variations today primarily reflect those present at birth, with the effects of radial redistribution generally adding only secondarily; in the most extreme case they double the pre-existing azimuthal scatter (a difference of $0.042$ dex for $3$ Gyr old stars).

For context, the orange line in Figure~\ref{fig:AZIM_Scatter_vs_Age_PUBLISH} also shows the total scatter (standard deviation) in [Fe/H] across each \textit{entire} galaxy ($R < 20 \kpc$).
In this case, there is no distinction between stars today versus at formation, given our in-situ selection criteria from Section~\ref{sec:selection}.
The galaxy-wide scatter is the same as the azimuthal scatter at $R = 8 \kpc$ for old stars, because old stars had and have negligible radial gradients.
However, rather than remaining constant for stars $\lesssim 8 \Gyr$ old, the galaxy-wide scatter rises toward younger ages, from the onset of the radial gradient, as \citet{Bellardini2022} showed.

\subsection{Comparison to the Milky Way}
\label{sec:Azimuthal_Comparison_to_the_Milky_Way}

\begin{figure*}
\centering
\includegraphics[scale = 0.52]{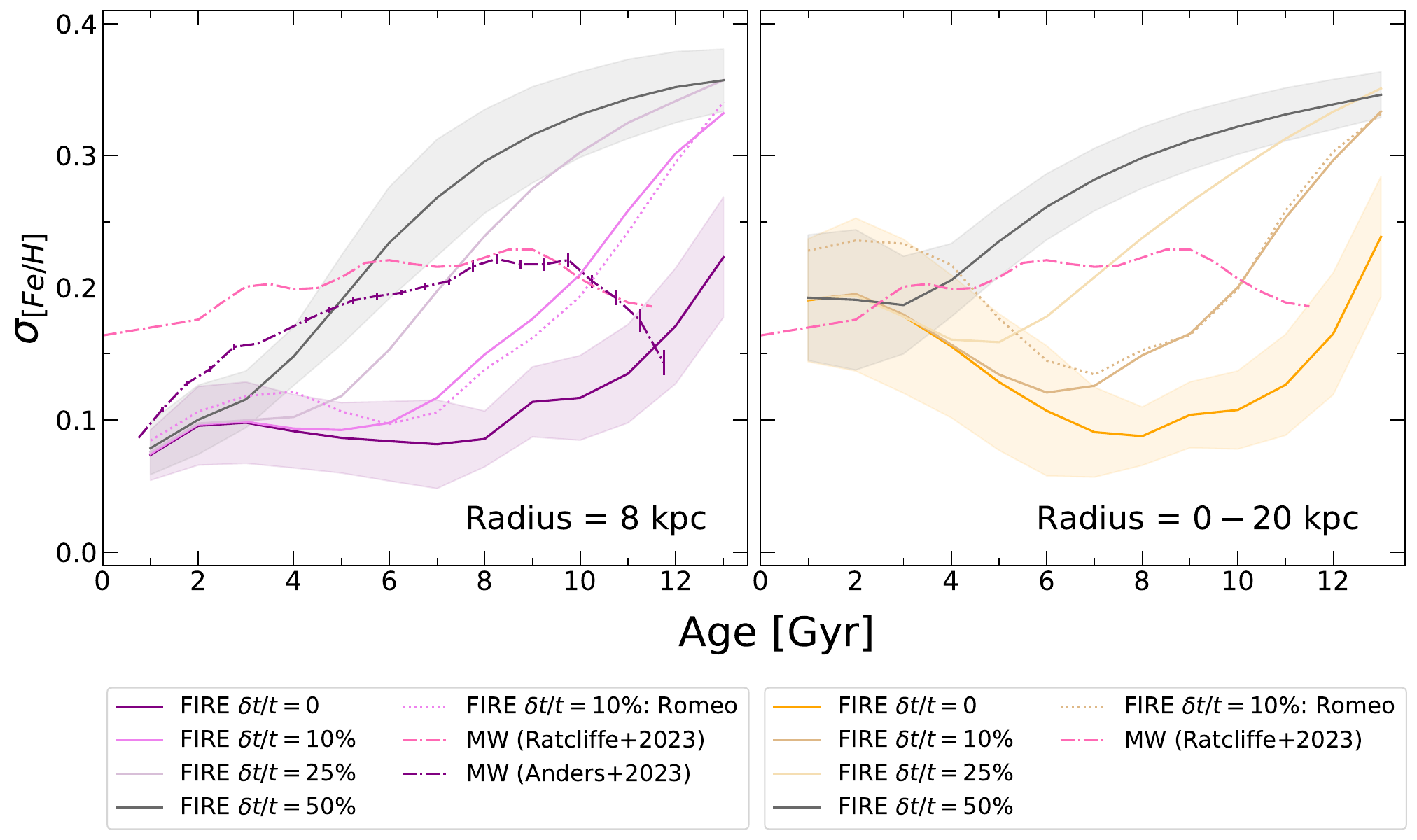}
\caption{Same as Figure~\ref{fig:AZIM_Scatter_vs_Age_PUBLISH}, but assuming various fractional uncertainties in stellar ages, for stars today at $R = 8 \pm 0.5 \kpc$ (left) and across the entire galaxy, $R = 0 - 20$ kpc (right).
Larger age uncertainty causes $\sigma_{\rm [Fe/H]}$ to increase more rapidly with age, leading to larger scatter at a given age.
For scatter across the entire galaxy (right), large age uncertainties can shift or erase the intrinsic minimum $\sigma_{\rm [Fe/H]}$ at ages $\approx 8 \Gyr$.
We also show measurements of the scatter across the MW from \citet{Anders2023} and \citet{Ratcliffe2023}.
Using a larger radial range and larger age uncertainty leads to a normalization in $\sigma_{\rm [Fe/H]}$ that is more similar with \citet{Ratcliffe2023}, especially at younger ages, but it does not match the relative flatness with age that \citet{Ratcliffe2023} finds.
FIRE-2 galaxies also do not match the downturn at ages $\gtrsim 9 \Gyr$ that both \citet{Anders2023} and \citet{Ratcliffe2023} find.
}
\label{fig:AZIM_Ratcliffe_Compare_2plots_PUBLISH}
\end{figure*}

Measuring the azimuthal scatter of metallicity in the MW is challenging. 
Two recent observational analyses of the MW presented results against which we compare, but we caution that, for azimuthal scatter in particular, modeling/matching the selection function is particularly important, and we defer a more careful comparison for future work.
For now, we compare with an emphasis on general trends.

\citet{Ratcliffe2023} used APOGEE DR17 to study elemental abundances in the MW through the derivation of birth radii of 145,447 red giant stars in the disk. Their Figure~2 quantifies the standard deviation in [Fe/H] (azimuthal scatter in the Solar neighborhood) versus stellar age, which we show in Figure~\ref{fig:AZIM_Ratcliffe_Compare_2plots_PUBLISH} (left and right).
Figure~\ref{fig:AZIM_Ratcliffe_Compare_2plots_PUBLISH} (left) also shows $\sigma_{\rm [Fe/H]}$ versus stellar age from Figure~12 of \citet{Anders2023}, also based on APOGEE (see Section~\ref{sec:radial_comparison}).
One important difference between these observational analyses is the effective radial range over which they measured variations.
\citet{Anders2023} sought to compute azimuthal scatter at a fixed $R$, that is, the scatter when subtracting out the effect of the radial gradient.
By contrast, \citet{Ratcliffe2023} computed the scatter across their sample, which encompasses a much larger radial range ($R \approx 5 - 12 \kpc$).
Therefore, for younger stars with ages $\lesssim 8 \Gyr$, the measured scatter is larger in \citet{Ratcliffe2023} than in \citet{Anders2023}, a difference that diverges towards younger ages.
Also, as a result, the age dependence of the scatter in \citet{Ratcliffe2023} is modest, varying from $\approx 0.16$ to $\approx 0.23$ dex across the entire age range.
Both works agree, though, that the scatter in [Fe/H] was largest $\approx 8 - 10 \Gyr$ ago.
Also, all observational works in Section~\ref{sec:radial_comparison} agree that the radial gradient in the MW is shallower for older stars, so if the radial gradient is a key contributor to the measured scatter in \citet{Ratcliffe2023}, this naively seems difficult to reconcile with the gradually rising trend with age in \citet{Ratcliffe2023}.

Figure~\ref{fig:AZIM_Ratcliffe_Compare_2plots_PUBLISH} compares these observational analyses to our FIRE-2 galaxies, for which we measure $\sigma_{\rm [Fe/H]}$ today versus stellar age.
We show results from FIRE-2 using two radial selections, to bracket the radial selections of \citet{Anders2023} and \citet{Ratcliffe2023}.
Specifically, the left panel shows stars at $R = 8 \pm 0.5 \kpc$ today, which in principle is most comparable to the analysis of \citet{Anders2023}.
Mimicking the selection function of \citet{Ratcliffe2023} is less straightforward, so we simply show the scatter across the entire galaxy ($R = 0 - 20 \kpc$) in the right panel, to conservatively bracket the extremes of radial selection.

We again assume various fractional uncertainties in stellar age, $\delta t / t$, in FIRE-2 when comparing with the MW.
Again, the effects of these age uncertainties are modest for $\delta t / t \lesssim 10\%$ and do not qualitatively change the trends, except perhaps for the oldest stars.
However, $\delta t / t \gtrsim 25\%$ affects the trend dramatically.
For the azimuthal scatter at $R = 8 \pm 0.5 \kpc$ (left), larger age uncertainties cause the scatter to rise much more rapidly with age at ages $\gtrsim 3 \Gyr$.
For the galaxy-wide scatter (right), larger age uncertainties cause the inferred minimum scatter to occur both at younger age and at larger scatter, thus ``flattening" the relation with age.
This highlights that inferred trends of azimuthal or galaxy-wide scatter with age are sensitive to stellar age uncertainties.

For young stars, with ages $\lesssim 1 \Gyr$, the azimuthal scatter in FIRE-2 at fixed $R$ (left) is comparable to that measured in \citet{Anders2023}.
FIRE-2 also qualitatively agrees with \citet{Anders2023} that the azimuthal scatter at fixed $R$ increases with age, if we include significant age uncertainties in FIRE-2.
However, the azimuthal scatter in \citet{Anders2023} initially rises more rapidly with age than in FIRE-2, and it also reaches a plateau at $\approx 0.22$ dex, unlike the FIRE-2 galaxies which continue to rise with age to $\approx 0.35$ dex (depending on age uncertainty).

The galaxy-wide scatter in FIRE-2 (right) is comparable to \citet{Ratcliffe2023} at ages $\lesssim 4$ Gyr.
This likely reflects a trade-off that the radial selection in \citet{Ratcliffe2023} is not quite as large as $20 \kpc$ but also that the MW has a stronger radial gradient than these FIRE-2 galaxies.
The behavior of the scatter with age in FIRE-2 in the right panel depends sensitively on stellar age uncertainties.
That said, none of the radial selections or age uncertainties applied to FIRE-2 agree with the relative flatness of the relation between scatter and age in \citet{Ratcliffe2023}.

We conclude that, while the FIRE-2 simulations show qualitatively similar trends to the MW for the radial and vertical gradients, these trends of the (azimuthal) scatter with age are qualitatively discrepant with those measured in the MW, despite that they agree reasonably well for just young stars.
In general, FIRE-2 shows azimuthal scatter that increases faster and monotonically with stellar age, as measured today with non-trivial age uncertainties, and no level of assumed age uncertainty matches the observed plateau at $\approx 9 \Gyr$.
Again, we caution that a more rigorous modeling of observational selection functions is essential for a more quantitative comparison.
We defer a more detailed analysis for future work.

\subsection{Relation between the radial gradient and azimuthal scatter}
\label{sec:relation}

As discussed in Section~\ref{sec:Azimuthal_FIRE}, the azimuthal scatter in these FIRE-2 galaxies is relatively independent of $R$ and is nearly independent of age at $\lesssim 8 \Gyr$. Here, we seek to understand the origin of this azimuthal scatter, at least for young stars today.
In particular, we test for its possible relation to the radial gradient.

We extend the results of \citet{Orr2023}, which studied the azimuthal variations of metals in gas in some of these same FIRE-2 galaxies at $z \approx 0$.
They found that radial transport of gas along spiral arms can be a significant driver for these azimuthal variations. In particular, spiral arms can act as ``metal freeways": their gravitational perturbations drive higher-metallicity gas from smaller radii outward and lower-metallicity gas from larger radii inward, thus contributing to the azimuthal scatter at a given $R$ today. \citet{Grand2016} previously demonstrated a similar effect using the Auriga cosmological zoom-in simulations, and more recently, using a (non-cosmological) $N$-body baryonic simulation, \citet{Khoperskov2022} found a similar dynamical phenomenon and concluded that radial gradients are the key ingredient for sourcing azimuthal variations across spiral arms.

As a test of the importance of this effect across a population, we examine whether galaxies with stronger radial gradients also have greater azimuthal scatter in their stellar metallicities, both at birth (as inherited from the star-forming gas) and today.

\begin{figure}
\centering
\includegraphics[width = \columnwidth]{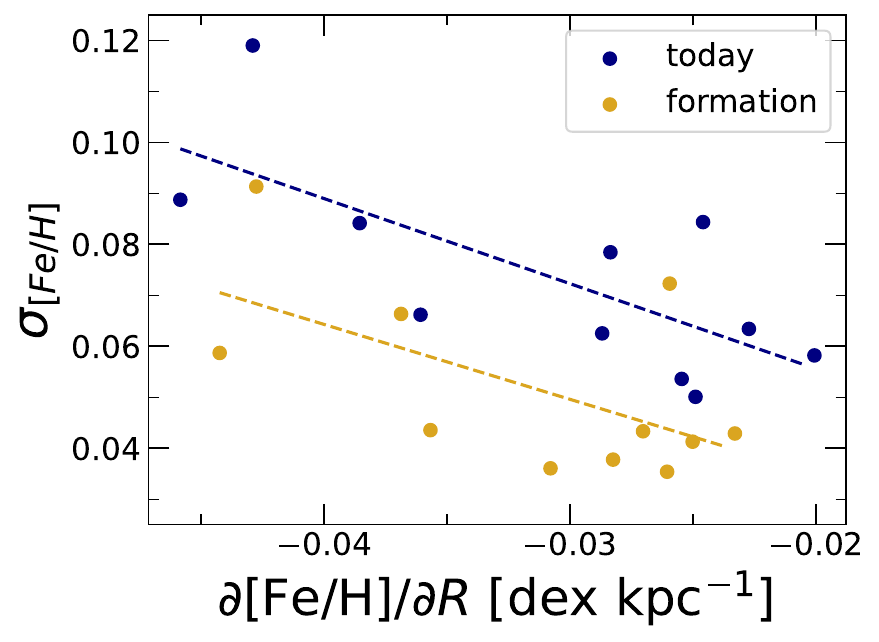}
\caption{Azimuthal scatter, $\sigma_{\rm [Fe/H]}$, at $R = 8 \pm 0.5 \kpc$ versus radial gradient, $\partial {\rm [Fe/H]} / \partial R$ across $R = 4 - 12 \kpc$, for stars younger than $1 \Gyr$ today, measured using radii today and at birth, for each of our 11 galaxies. Galaxies with larger azimuthal scatter have stronger radial gradients, with best-fit slopes of $-1.7$ kpc today (navy dashed line) and $-1.5$ kpc at formation (gold dashed line).
This super-linear relationship supports the idea that the azimuthal scatter at late cosmic times (once the disk has settled) is primarily sourced by the radial redistribution of gas before star formation (as the trend at formation shows) and additionally by the radial redistribution of stars after formation (as the trend today shows), which mixes the radial gradient into azimuthal scatter.
} 
\label{fig:RAD_slope_vs_AZIM_scatter_PUBLISH}
\end{figure}

Figure~\ref{fig:RAD_slope_vs_AZIM_scatter_PUBLISH} shows the azimuthal scatter in stars, $\sigma_{\rm [Fe/H]}$, at $R = 8 \pm 0.5 \kpc$ versus the radial gradient, $\partial {\rm [Fe/H]} / \partial R$, within $\pm 4 \kpc$ of $R = 8 \kpc$, for stars with age $< 1 \Gyr$ both today and at birth.
We use $\pm 4 \kpc$ as a conservative upper limit to the amount of redistribution from their birth radius that stars experience over $\approx 1 \Gyr$ in FIRE-2 (Bellardini et al. (in prep.). Each point shows one of the 11 FIRE-2 galaxies, which we measure both today and using the coordinates of the stars at birth. Galaxies with larger azimuthal scatter have a stronger radial gradient. We test the strength of this correlation using a Pearson test, finding a coefficient of $-0.72$ and p-value of $0.012$ for stars today, and a somewhat weaker coefficient of $-0.59$ and p-value of $0.054$ for stars at formation.
Therefore, both correlations are relatively strong and significant.

Figure~\ref{fig:RAD_slope_vs_AZIM_scatter_PUBLISH} also shows a fit linear relation between $\sigma_{\rm [Fe/H]}$ and $\partial {\rm [Fe/H]} / \partial R$, which has a slope of $-1.7 \kpc$ for stars today and $-1.5 \kpc$ for stars at formation.
It is noteworthy that both relations are steeper than linear. As previously discussed, we necessarily measure the azimuthal scatter using a finite radial bin of $\pm 0.5 \kpc$, so a galaxy with a stronger radial gradient would have stronger (radial) variation within this bin just from the radial gradient.
However, as Figure~\ref{fig:RAD_slope_vs_AZIM_scatter_PUBLISH} shows, the azimuthal scatter is generally much stronger than the strength of the radial gradient multiplied by the radial bin width.
This means that the azimuthal scatter is not simply dominated by radial variations within the bin, suggesting other contributions. To further test this, we reduced the radial bin width in measuring the azimuthal scatter from $8 \pm 0.5 \kpc$ to $8 \pm 0.25 \kpc$: the scatter was essentially unaffected ($\Delta \lesssim 0.001$ dex/kpc).
Thus, the scatter that we present represents the true azimuthal variations along an annulus.
%Therefore, for radial annuli less than $1 \kpc$ in width at the solar annulus, azimuthal variations dominate over variations from the radial gradient, further confirming that radial redistribution of gas and stars into these regions are the source of these variations.

This relation between azimuthal scatter and the radial gradient supports the conclusions of \citet{Orr2023} and \citet{Khoperskov2022} that spiral arms act as ``metal freeways", and that strong radial gradients are a key contributor to azimuthal variations in spiral arms.
This predicted correlation would be compelling to test using a large sample of observed disk galaxies.

\section{Summary and Discussion}
\label{sec:discussion}

\subsection{Summary}
\label{sec:summary}

We analyzed 11 MW-mass galaxies from the FIRE-2 suite of cosmological zoom-in simulations.
As several works have shown, they reproduce many key observed properties of the MW, M31, and similar-mass galaxies in the nearby universe.
We analyzed the radial, vertical, and azimuthal variations of elemental abundances in stars.
We focused on [Fe/H], but in Appendix~\ref{sec:appendix1} we also show trends for [Mg/H] and [C/H], which are qualitatively similar.
These simulations allow us to examine these trends both today and at the time that stars formed, in a realistic cosmological setting.
The key question that we explored is the extent to which spatial variations in elemental abundances today reflect those at birth.
We also compared with the MW, including effects of uncertainties in stellar age in analyzing the simulations.
Our main results are as follows.

%\begin{itemize}

\textit{All spatial variations today (radial, vertical, azimuthal), including their trends with age and location, were already in place at the time that stars formed.}
The evolution of stars after birth, via radial redistribution and vertical heating, plays a subdominant role in determining spatial patterns in elemental abundances today.
That said, the combination of radial redistribution and vertical heating of stars has somewhat weakened (washed out) the radial and vertical gradients, and somewhat increased the azimuthal scatter, since birth.
%At the time of formation, radial and vertical gradients were systematically stronger, and azimuthal scatter was systematically weaker, than they are today.
%However, these differences are generally small compared with the trends already in place at birth.

\textit{Radial gradients}:
In FIRE-2 they are flat for the oldest ($\gtrsim 12 \Gyr$) stars and steepen systematically for younger stars.
Importantly, this is true both today and at formation, because the metallicity radial gradient in the ISM became steeper (more negative) over time.
Radial gradients in the MW are generally stronger than in FIRE-2 galaxies at most ages today.
We speculate that this is largely because the MW's disk settled earlier than the galaxies in our FIRE-2 sample.
That said, both FIRE-2 and the MW show the same key trend of a steeper radial gradient for younger stars, and both show nearly flat radial gradients when measuring all stars (regardless of age) today.
This implies that the radial gradient in the MW today largely reflects what was in place in the MW at birth: the ISM had progressively steeper (more negative) radial gradients in metallicity over time.

\textit{Vertical gradients}:
In FIRE-2 they have the opposite behavior as the radial gradients.
The vertical gradient for all stars today is strong, while the vertical gradient for mono-age stars is mostly negligible.
The latter is likely because the vertical extent of the star-forming gas disk is set by turbulence, which also efficiently mixes metals.
Therefore, the strong vertical gradient we find when considering stars of all ages, present both today and at birth, primarily reflects the upside-down settling of the disk over time, while it enriched in metals.
These vertical gradients are strongest at small radii, where the range of stellar ages is largest, and they flatten with radius, being minimal in the outer disk where the age range is minimal.
Vertical gradients of all stars today are shallower in FIRE-2 galaxies than in the MW at most radii, including at the solar annulus ($R = 8 \kpc$). We speculate that the MW's disk settled earlier and/or enriched more rapidly in metals as it settled than in these FIRE-2 galaxies.
However, the general trend of a steeper vertical gradient at smaller radii agrees with the MW.

\textit{Azimuthal scatter (at fixed radius)}:
In FIRE-2 it is nearly independent of radius today, and it increased weakly with radius at birth.
At $R = 8 \kpc$ (Solar annulus), the azimuthal scatter is nearly constant with stellar age over the last $\approx 8 \Gyr$ at $\sigma_{\rm [Fe/H]} \approx 0.087$ dex today and $\approx 0.062$ dex at birth after the galactic disk settled, and it rapidly rises with age before that from the increasing ``patchiness'' of the gas at earlier times.
The trend of azimuthal scatter versus age is different in FIRE-2 than in the MW, with the MW showing an increase with age at young ages, a peak at $\approx 9$ Gyr, and then a decrease with age toward older ages.

\textit{Effects of stellar age uncertainties}:
Adding fractional uncertainties in stellar age of $\delta t / t \gtrsim 25\%$ to these FIRE-2 galaxies can dramatically change (bias) the age dependence of these spatial variations.

\textit{Relation between radial gradients and azimuthal scatter}:
In FIRE-2, galaxies with larger azimuthal scatter in metallicity for young stars today have a stronger radial gradient.
This implies that azimuthal scatter today is largely a result of radial redistribution of gas and stars, likely driven by spiral arms \citep[following][]{Orr2023}.

\textit{Conclusion}:
While the FIRE-2 simulations generally show weaker gradients than the MW, the key trends with stellar age and location today are similar. Because spatial variations in metallicity of stars today in FIRE-2 were largely in place when the stars formed, this therefore implies that the spatial variations in metallicity of stars in the MW today were also largely in place when the stars formed.

\subsection{Caveats}
\label{sec:caveats}

We describe some important caveats to our analysis.
First, we analyzed a sample of 11 MW-mass galaxies.
Their cosmological selection function was simple: isolated or LG-like paired dark-matter halos at $z = 0$ with $M_{\rm 200m} \approx 1 - 2 \times 10^{12} \Msun$, with no additional selection based on galaxy properties or merger/formation history.
Our results thus sample random formation histories of MW-mass galaxies.
Indeed, as \citet{Bellardini2021} showed, the radial gradients and azimuthal variations of gas metallicity in these simulations at $z \approx 0$ are comparable to, and in some cases steeper than, those measured in M31 and in surveys of nearby MW-mass galaxies.
That said, as our comparisons have shown, these FIRE-2 simulations are not perfect analogs of the MW; in particular, they generally have shallower gradients.

Furthermore, these FIRE-2 simulations do not model potentially important physical ingredients, including magnetohydrodynamics or self-consistent cosmic rays injection and transport \citep[see discussions in][]{Hopkins2018, Wetzel2023}.
However, analyses of FIRE-2 simulations that included these processes indicate that they are not likely to affect disk dynamics (and therefore disk-wide metallicity variations) \textit{at fixed stellar mass} \citep[for example,][]{Su2017, McCluskey2023}.
Potentially more important is that these FIRE-2 simulations do not include supermassive black holes and their AGN feedback.
While the effects of AGN feedback are likely most significant in the innermost regions of galaxies, and less likely to impact stellar dynamics and star-formation rates in the Solar annuli of MW-mass galaxies \citep{MercedesFeliz2023, Wellons2023}, some simulations  \citep[for example][]{Irodotou2022} indicate that AGN feedback can affect the kinematic and structural properties of MW-mass galaxies.

Finally, in terms of stellar nucleosynthesis in FIRE-2, all results are based on a single set of models, which are themselves uncertain.
For example, \citet{Gandhi2022} showed that the implementation of white-dwarf supernovae in FIRE-2 likely underestimates their rates, leading to a modest underproduction of [Fe/H].
As we show in Appendix~\ref{sec:appendix1}, our key results are largely insensitive to which elements we examine, despite their contributions from different stellar evolution channels, which suggests that the trends that we showed are likely robust to current uncertainties in stellar nucleosynthesis.
Another key limitation, especially for our results on azimuthal variations, is that FIRE-2 models all core-collapse supernovae via the same (IMF)-averaged yields, but \citet{Muley2021} showed how different mass progenitors lead to different patterns in abundances, in particular, with larger scatter.

\subsection{The strength of gradients compared with the Milky Way}
\label{sub:stength_of_rad_grad}

As discussed in Section~\ref{sec:radial_comparison}, and in detail in \citet{Bellardini2021}, the radial gradients of metallicity in these FIRE-2 MW-mass galaxies generally agree well with measurements of M31, and they are generally \textit{steeper} than measurements of populations of nearby MW-mass galaxies.
However, they also are generally shallower than in the MW: the strongest gradient in FIRE-2 agrees with the shallowest gradient measured in the MW, but otherwise observations of the MW are consistently steeper.
At face value, these comparisons with nearby galaxies suggest that the MW has an unusually steep gas-phase radial metallicity gradient \citep[see also][]{Boardman2020}.

One contribution to this could be distance uncertainties in measurements of the MW. \citet{Donor2018} found that the distance catalog used can change observed radial gradients by up to $40\%$.

Another possibility is that the MW disk settled unusually early compared to our simulations and compared with nearby galaxies, as \citet{Bellardini2022} argued, finding a weak, though non-zero, correlation between the strength of the radial gradient and the time of disk settling.
\citet{McCluskey2023} found that the kinematics for young stars in FIRE-2 agree reasonably well with the MW and nearby galaxies, but McCluskey et al. (in prep.) compares these kinematics as a function of stellar age against the MW, to understand better how the dynamical evolution of these galaxies compares with the MW \citep[see also][]{Belokurov2022}.

\subsection{Comparison to previous works}
\label{sub:Compare_to_other_cos_sim}

Our work builds on previous analyses of the FIRE cosmological zoom-in simulations \citep{Ma2017a, Ma2017b, Bellardini2021, Bellardini2022}, all of which agree with the trends we presented.
Various other works using different cosmological simulations also studied elemental abundance patterns.

\citet{Font2011} analyzed the trends in the radial profiles of [Fe/H] in stars today using the GIMIC suite of cosmological simulations of 412 galaxies. They found a typical radial gradient for all stars measured across $R < 20 \kpc$ of $\partial {\rm [Fe/H]} / \partial R \approx -0.025$ dex kpc$^{-1}$. More recently, \citet{Font2020} used the ARTEMIS simulations to study the [Fe/H] profile in 42 MW-mass galaxies and found a gradient for all stars measured across the same range of $\approx -0.015$ dex kpc$^{-1}$. While steeper than the profiles we show in Figure~\ref{fig:RAD_profile_allstars_AND_RAD_normalized_profile_2plots_PUBLISH}, both works agree with our negative radial gradient in [Fe/H] for all stars at $R \gtrsim 6 \kpc$.

Several works also studied how the radial profile of metallicity changes over time in the (star-forming) ISM and/or in stars at birth. 
Some works found that radial gradients were steeper in the past and became shallower (flatter) over time, opposite to our results.
\citet{VK2018} analyzed [O/H] profiles in the ISM for 10 MW-mass galaxies in their cosmological simulations and found gradients measured across $R \approx 0 - 15 \kpc$ to be strongest at early times. \citet{Agertz2021} studied the [Fe/H] profile in the ISM using the VINTERGARTEN cosmological simulation of a single MW-mass galaxy and found similar results, though they note that a relatively late-time merger may have affected the evolution of this particular galaxy.
\citet{Buck2023} analyzed 4 galaxies from the NIHAO-UHD suite of cosmological simulations and found steeper gradients in gas at earlier times, and, in particular, that the gradients only steepened shortly after a gas-rich merger, which were more common in the past.
However, they measured such gas-phase gradients across a fixed radial range, $R = 2.5 - 17.5 \kpc$, across cosmic time, which differs significantly from our measuring such gradients only out to radii where stars form, $\approx R_{\rm 90}$, which may contribute to these different trends.

Other analyses of cosmological simulations found results similar to ours.
\citet{VK2020} studied the [Fe/H] profile of stars at birth using a cosmological zoom-in simulation of a single MW-mass galaxy and found a steepening gradient at later times.
Similarly, \citet{Lu2022b} analyzed 4 simulations from the NIHAO-UHD suite and found only mild evolution in the [Fe/H] radial gradient of the ISM over time when measured over $R = 0 - 10 \kpc$, with some galaxies showing mild steepening and others showing mild shallowing over time. 

Therefore, expectations from cosmological simulations are mixed regarding the strength and even direction of the evolution of radial gradients in the ISM across the histories of MW-mass galaxies.
The details of the modeling of star formation and feedback can lead to differing expectations.
For example, \citet{Pilkington2012} and \citet{Gibson2013} analyzed the MUGS and MAGICC suites of cosmological simulations. The ``weaker'' feedback model of MUGS results in stronger radial [O/H] gradients in gas at earlier times, which flatten over time. The ``enhanced" feedback model of MAGICC (otherwise identical to MUGS) redistributes and mixes the ISM and washes out any gradients early on, resulting in flat profiles at early times that slightly steepen over time, as the rotation of the disk becomes well-ordered.

Analyzing the larger-volume TNG50 simulation, 
\citet{Hemler2021} studied $1595$ galaxies with $M_{\rm star} = 10^9 - 10^{11} M_\odot$, and found radial gradients in gas-phase [O/H] to be steeper in galaxies at higher redshift. On the other hand, \citet{Tissera2022} studied $957$ galaxies with $M_{\rm star} > 10^9 M_\odot$ from the EAGLE simulation and found radial gradients in gas-phase [O/H] to have essentially no evolution ($\approx 0.001$ dex kpc$^{-1}$ $/ \delta z$) with redshift.
We note, however, that \citet{Hemler2021} and \citet{Tissera2022} tracked galaxies of the same mass over cosmic time, rather than the histories of individual galaxies as we did, which complicates the comparison, given expectations (at least for FIRE) that the strength of the radial gradient correlates with galaxy mass at a given redshift \citep[for example,][]{Ma2017a}. 

Fewer works have studied vertical profiles of metallicity using simulations or analytical models. \citet{Minchev2013} used a cosmological simulation of a single MW-mass galaxy to study the vertical profile of stars at $R = 8 \pm 1 \kpc$. They found that [Fe/H] decreases with $|Z|$ when considering all stars, which agrees with our Figure~\ref{fig:VER_profile_allstars_PUBLISH}, and that mono-age vertical gradients are essentially flat, which agrees with our Appendix~\ref{sec:appendix1}.

Regarding azimuthal scatter, \citet{Lu2022b} studied such variations in the ISM in 4 MW-mass galaxies from NIHAO-UHD and found that $\sigma_{\rm [Fe/H]}$ slightly increases with radius, in agreement with our Figure~\ref{fig:AZIM_profiles_4plots_PUBLISH}. Furthermore, \citet{Lu2022b} found a generally increasing $\sigma_{\rm [Fe/H]}$ with increasing lookback time, also agreeing with our Figure~\ref{fig:AZIM_Scatter_vs_Age_PUBLISH}. 
\citet{Khoperskov2022} used 3 controlled (non-cosmological) $N$-body baryonic simulations to uncover the origin of abundance variations across spiral arms in MW-like disk galaxies, and found that radial gradients play a significant role in sourcing azimuthal variations across spiral arms, in agreement with our Figure~\ref{fig:RAD_slope_vs_AZIM_scatter_PUBLISH}.

Analytic models also explored the expected evolution of spatial patterns in abundance.
\citet{Minchev2018} developed a method for determining stellar birth radii in the MW by extrapolating the evolution of the ISM metallicity with radius and time. They argued for the expectation that the radial gradient became shallower over time, from the ``inside-out'' nature of disk growth, that is, increasing fractional stellar mass grows at increasing radii over time.
They also argued that radial redistribution plays a more significant role in flattening radial gradients for older stars, to become as weak as they are in the MW today. 
While we agree with \citet{Minchev2018} that radial redistribution acts to flatten radial gradients in mono-age stellar populations, we disagree with the expectation that the radial gradient in the ISM was steeper in the past.
The FIRE simulations show the opposite trend, despite also experiencing radial inside-out formation, at least on average.
In earlier work, \citet{Chiappini2001} developed an elemental evolution model and argued that radial gradients in metallicity steepen over time, in agreement with our results.
They explained this as occurring from infalling metal-poor gas in the outer galaxy at later times, which differs from the reasons for the steepening gradient over time that we argue for.
More recently, \citet{Sharda2021} developed an analytic framework that argued for a steepening of radial gradients over time.

Observationally, \citet{Cheng2024} studied [O/H] gas-phase radial gradients using a sample of 238 star-forming galaxies at $z = 0.6 - 2.6$ and found that galaxies have generally flat, and sometimes \textit{positive} radial gradients, at these early times, which tend to steepen in a negative direction over time.

\subsection{Evolution of radial gradients}
\label{sec:Radial_Gradients_at_Formation}

The most important source of spatial variations in metallicity for a disk galaxy like the MW, at least today, is the radial gradient.
And yet, as we described above, theoretical works continue to debate the expected strength and even simply the sign of the evolution of the radial gradient over a galaxy's history.
Determining the radial gradient of the ISM of the MW across its history, which set the birth conditions of stars, has been challenging via galactic archeology, given additional effects like radial redistribution of stars after birth that complicate interpreting their patterns today.
However, as we showed in Section~\ref{sec:radial_comparison}, essentially all recent analyses of the MW agree that the radial gradient today is systematically shallower (less negative) for older stars.
The simplest explanation for this observed trend is that it simply reflects the conditions of the (star-forming) ISM across the MW's history, such that the radial gradient of ISM metallicity in the MW became steeper (more negative) over time, and that the effects of stellar radial redistribution were not so strong as to reverse or undo this trend.

We showed that this ``simple" behavior indeed occurs in the FIRE cosmological simulations.
We reiterate arguments as to why this behavior should emerge, as previous works articulated \citep[for example,][]{Ho2015, Ma2017a, Bellardini2021}.

First, to \textit{source} a metallicity radial gradient (in principle), smaller radii must have higher star-formation efficiency, that is, mass in (young) stars that source most of the metals, relative to mass in gas, over which the metals are diluted.
In other words, $\log(\Sigma_{\rm star}(R) / \Sigma_{\rm gas}(R)) \propto {\rm [Fe/H]}(R)$ (at least for some age stellar population) should decline with $R$.
Gas generally distributes within a galaxy with a declining (surface) density profile, often as $\Sigma_{\rm gas} \sim \exp(-R / R_{\rm s})$, where $R_{\rm s}$ is the scale radius.
Furthermore, star formation density generally exhibits a super-linear relation to total gas density \citep{Schmidt1959, Kennicutt1998}, such that $\Sigma_{\rm SFR} \propto \Sigma_{\rm gas}^{n}$, often with $n \approx 1.4$.
As long as $n > 1$, this first condition should be met, modulo the complications arising from metal mixing and metal outflows.
Indeed, \citet{Bellardini2022} showed that in these FIRE-2 MW-mass simulations at $z \approx 0$, $\log(\Sigma_{\rm star}(< 1 \Gyr, R) / \Sigma_{\rm gas}(R)) \propto {\rm [Fe/H](R)}$ to good approximation.

Second, to \textit{sustain} a radial gradient in gas, the galaxy must have sufficiently well-ordered rotation (in a disk) to prevent radial mixing of metals via turbulence. As these FIRE-2 MW-mass galaxies evolved, their ISM developed thinner disks with more ordered rotation \citep{McCluskey2023}, thereby reducing radial mixing, which is a key reason why the radial gradients became steeper over time.
Indeed, for young stars in these FIRE-2 galaxies, \citet{Bellardini2022} showed a significant correlation between the strength of the radial metallicity gradients and the ``diskiness" of the galaxy, defined by $v_{\rm \phi} / \sigma_{\rm v}$.
More broadly, \citet{Ma2017a} analyzed the FIRE-1 simulations across a broad range of galaxy masses and redshifts and found that only galaxies with $v_{\rm \phi} / \sigma_{\rm v} \gtrsim 1$ showed significant (negative) radial gradients.

This idea of higher star formation efficiency at smaller radii in conjunction with more ordered rotation leading to a \textit{strengthening} of radial gradients agrees with \citet{Khoperskov2022}, which used controlled (non-cosmological) $N$-body baryonic simulations of disk galaxies to show that their galaxy that experienced enrichment over time developed a stronger radial gradient than their (otherwise identical) galaxy that did not.

We emphasize that this steepening of radial metallicity gradients in the FIRE cosmological simulations occurs even though they (on average) experienced ``inside-out" radial growth in stars, that is, higher fractional growth rate in stars at larger radii at later times.
This contrasts with arguments, such as in \citet{Minchev2018}, that inside-out radial growth should lead to ISM radial gradients becoming shallower over time.

Of course, we presented average/typical trends across 11 MW-mass FIRE-2 galaxies.
Individual galaxies, including the MW, could exhibit atypical behavior, and the evolution of quantities like radial gradients can be non-monotonic, especially from events like galaxy mergers \citep[for example,][]{Ma2017a, Buck2023}. More measurements of stellar populations across the MW, with increased accuracy in determining stellar ages and birth radii, are necessary to address this question for the MW's history specifically.
Nonetheless, we reiterate that the trends in the FIRE-2 cosmological simulations are similar to the trends observed for the MW, which provides evidence that they do reflect, to a significant degree, the history of the MW.

\section{Acknowledgements}

We thank Fiona McCluskey for detailed comments.
RG, AW, and MB received support from: NSF via CAREER award AST-2045928 and grant AST-2107772; NASA ATP grant 80NSSC20K0513; HST grants AR-15809, GO-15902, GO-16273 from STScI.

We ran simulations using: XSEDE, supported by NSF grant ACI-1548562; Blue Waters, supported by the NSF; Frontera allocations AST21010 and AST20016, supported by the NSF and TACC; Pleiades, via the NASA HEC program through the NAS Division at Ames Research Center.

The data in these figures, and the Python code that we used to generate them, are available at \url{https://github.com/rlgraf/Graf-et-al.-2024.git}.
FIRE-2 simulations are publicly available \citep{Wetzel2023} at \url{http://flathub.flatironinstitute.org/fire}.
Additional FIRE simulation data is available at \url{http://fire.northwestern.edu/data}.
A public version of the \textsc{Gizmo} code is available at \url{www.tapir.caltech.edu/~phopkins/Site/GIZMO.html}.

\software{GizmoAnalysis \citep{GizmoAnalysis}}

\bibliographystyle{aasjournal}
\bibliography{article}{}

\appendix
%\begin{multicols}{2}

\twocolumngrid
\counterwithin{figure}{section}

\section{Vertical gradients for mono-age stars}
\label{sec:appendix2}

\begin{figure*}[ht]
\centering
\includegraphics[scale = 0.42]{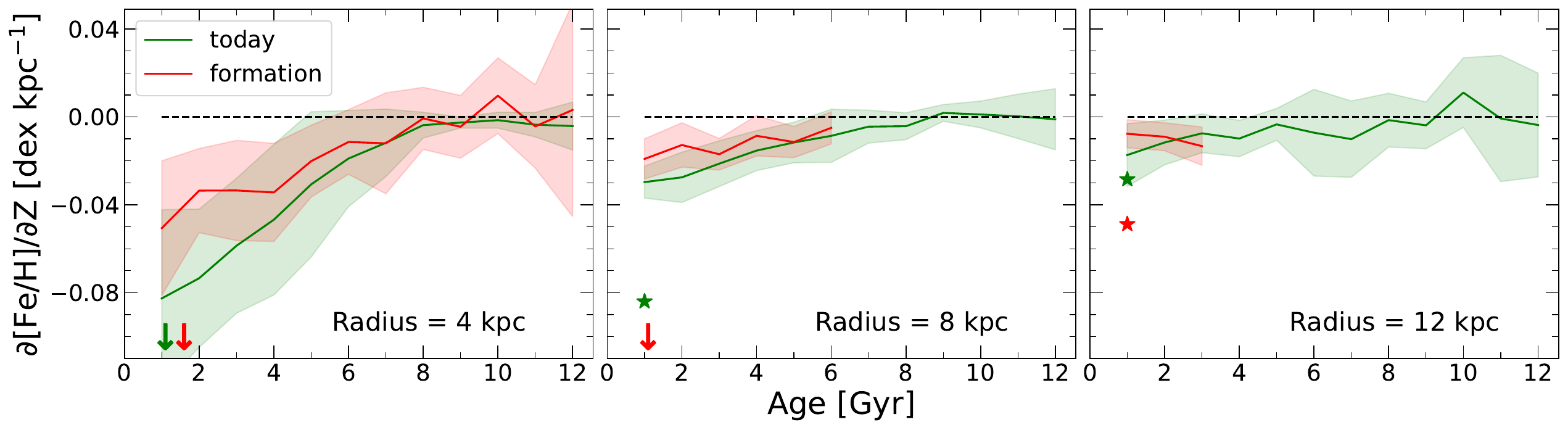}
\caption{Vertical gradient of [Fe/H], $\partial {\rm [Fe/H]} / \partial |Z|$, of stars across $|Z|< 1 \kpc$, versus age, for stars both today and at formation (using birth $|Z|$ and $R$), at different $R$ ($\pm 0.5$ kpc): 4 kpc (inner disk), 8 kpc (Solar annulus), and 12 kpc (outer disk).
Each line and shaded region show the average and $1 \sigma$ scatter across the 11 galaxies.
The vertical gradient for mono-age stars tends to be steeper for younger stars, both today and at formation, and the vertical gradient is generally steeper today than at formation, likely from the effects radial redistribution.
However, there is little-to-no vertical gradient for older stars or at larger $R$.
Most importantly, for comparison, the stars show the vertical gradient for all stars regardless of age from Figure~\ref{fig:VER_profile_allstars_PUBLISH}, and arrows indicate values that extend below plot, which are: $-0.32$ dex kpc$^{-1}$ (today) and $-0.46$ dex kpc$^{-1}$ (formation) at $4 \kpc$, and $-0.15$ dex kpc$^{-1}$ at formation at $8 \kpc$.
Thus, these vertical gradients for mono-age stars are always much smaller than the total gradient for all stars.
\textit{The vertical gradient of all stars today primarily reflects the ``upside-down'' settling of the disk over time, and not the gradient of mono-age stars.}
}
\label{fig:VER_slope_3plots_PUBLISH}
\end{figure*}

In Section~\ref{sec:Vertical_Profile} we showed the vertical profiles and gradients of all stars, regardless of age. Figure~\ref{fig:VER_slope_3plots_PUBLISH} shows the vertical gradients, $\partial {\rm [Fe/H]} / \partial |Z|$, for mono-age stellar populations versus their age, both today and using their coordinates at birth, at three radii ($\pm 0.5 \kpc$): $R = 4 \kpc$, $R = 8 \kpc$, and $R = 12 \kpc$. As in Figure~\ref{fig:VER_slope_vs_R_PUBLISH} (left), we measure the vertical gradient across $|Z| < 1 \kpc$.
For comparison, Figure~\ref{fig:VER_slope_3plots_PUBLISH} shows the vertical gradient of all stars at that $R$, denoted with a star, though in some cases the value extends well below the bottom axis, so we show an arrow.
The vertical gradient of mono-age stars is fairly flat at all ages, at least compared with the gradient for all stars.
This is because the vertical height of the disk is largely regulated by turbulence (the size of the largest turbulent Jeans length), and this same turbulence efficiently mixes metals vertically in the ISM, so stars forming at a given time reflect this weak vertical gradient.
This turbulent mixing is, in part, the same reason these galaxies do not have strong radial gradients for old stars.

That said, these FIRE-2 galaxies do show significant vertical gradients in the inner galaxy ($R \lesssim 4 \kpc$) for young stars ($\lesssim 6 \Gyr$), as \citet{Bellardini2022} (their Figure~8) already noted for stars at the time of formation.
As \citet{Bellardini2022} argued, likely this reflects the higher density and efficiency of star formation in the inner galaxy, such that in this regime the enrichment timescale can be shorter than the mixing timescale.
Furthermore, Figure~\ref{fig:VER_slope_3plots_PUBLISH} shows that, unlike vertical gradients for all stars in Figure~\ref{fig:VER_slope_vs_R_PUBLISH}, these mono-age gradients tend to be steeper today than at formation. 
Vertical heating alone would tend to drive this change in the opposite direction, so likely this arises from radial redistribution, potentially combined with disk flaring: higher-metallicity stars from the inner galaxy redistribute outward, enriching the galaxy at lower $|Z|$, and lower-metallicity stars from outward redistribute inward, lowering the metallicity at higher $|Z|$.

In summary, Figure~\ref{fig:VER_slope_3plots_PUBLISH} shows that the vertical gradient for mono-age stars is always significantly smaller than that for all stars, so the vertical gradient for all stars primarily reflects disk vertical settling over time, as opposed to a vertical gradient in star-forming gas.
This is the opposite behavior to the radial gradient, as in Figure~\ref{fig:RAD_profile_allstars_AND_RAD_normalized_profile_2plots_PUBLISH}, for which the radial gradient can be strong for star-forming gas, and thus for mono-age stars, but it is nearly flat when measuring all stars today.

\section{Spatial variations of different elements}
\label{sec:appendix1}

\begin{figure}
\centering
\includegraphics[width = \columnwidth]{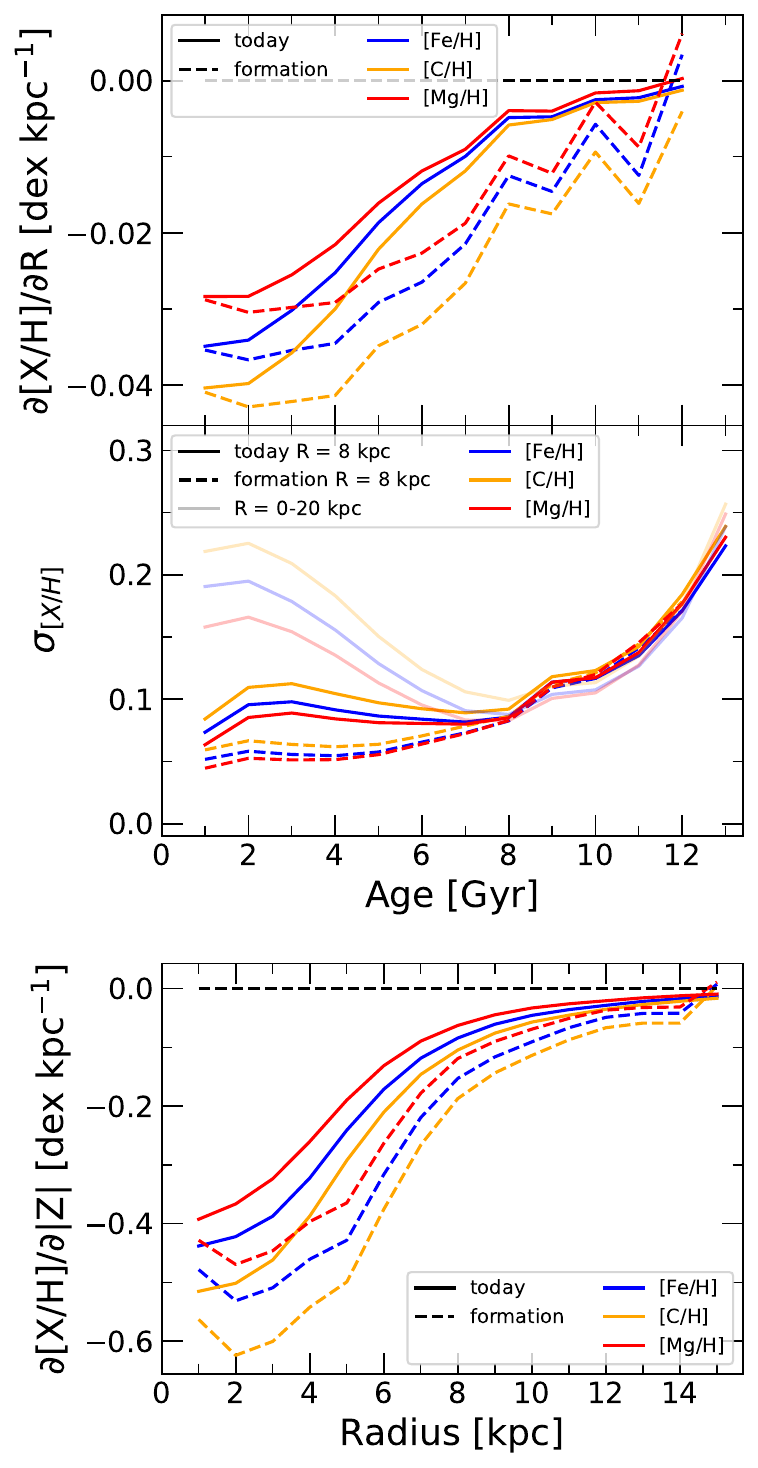}
\caption{Comparing the result for [Fe/H], [C/H], and [Mg/H], for the radial gradient, $\partial {\rm [X/H]} / \partial R$ (top) and the azimuthal/overall scatter, $\sigma_{\rm [X/H]}$, (middle), versus age, and the vertical gradient, $\partial {\rm [X/H]} / \partial |Z|$, versus cylindrical radius, $R$, (bottom), of stars both today and at formation (using their birth radii).
We use the same methods as in Figures~\ref{fig:RAD_slope_PUBLISH}, \ref{fig:AZIM_Scatter_vs_Age_PUBLISH}, and \ref{fig:VER_slope_vs_R_PUBLISH}. Radial and vertical gradients are somewhat steeper and the scatter is somewhat higher in [Fe/H] than in [Mg/H], given the significant contribution to [Fe/H] from white-dwarf (Ia) supernovae in the inner galaxy from older stars.
The gradients are somewhat steeper and the scatter is somewhat higher for [C/H], given its significant contribution from (progenitor-metallicity-dependent) stellar winds in FIRE-2.
\textit{However, the radial gradients, vertical gradients, and azimuthal scatter are all qualitatively similar for all 3 elements, both today and at birth.}
}
\label{fig:RAD_and_VER_slope_and_AZIM_Fe_H_C_Mg_3plots_PUBLISH}
\end{figure}

Throughout we used [Fe/H] as our representative measure of metallicity, as is common observationally.
However, as we discussed in Section~\ref{sec:Metal_enrichment_in_FIRE2}, FIRE-2 tracks 11 elements.
Here we present results for [C/H] and [Mg/H] as well.
These elements trace different stellar nucleosynthesis channels.
In the FIRE-2 model, [Fe/H] is sourced primarily from white-dwarf (Ia) supernovae (though with a significant contribution from core-collapse supernovae as well), [C/H] primarily through stellar winds, and [Mg/H] primarily through core-collapse supernovae.

Figure~\ref{fig:RAD_and_VER_slope_and_AZIM_Fe_H_C_Mg_3plots_PUBLISH} (top) shows the radial gradient, $\partial {\rm [X/H]} / \partial R$, versus stellar age for [Fe/H], [C/H], and [Mg/H], both today and at birth, using the same selection criteria as in Figure~\ref{fig:RAD_slope_PUBLISH}. Radial gradients for [Fe/H] are systematically stronger than for [Mg/H], because of the significant population of older stars in the inner galaxy, which leads to an enhanced rate of white-dwarf supernovae, and thus Fe enrichment, there. [C/H] shows the steepest gradients, given that its yields from stellar winds increase with progenitor metallicity in the FIRE-2 model. Despite these differences, the prevailing trends in the radial gradients of all three elements are similar, both today and at formation, in agreement with \citet{Bellardini2021} for gas and \citet{Bellardini2022} for stars at formation.

Figure~\ref{fig:RAD_and_VER_slope_and_AZIM_Fe_H_C_Mg_3plots_PUBLISH} (middle) compares the trends for azimuthal variations, $\sigma_{\rm [X/H]}$, versus stellar age, following the methods of Figure~\ref{fig:AZIM_Scatter_vs_Age_PUBLISH}. As expected, given our results in Section~\ref{sec:relation}, the element with the weakest radial gradient, [Mg/H], also has the lowest azimuthal scatter, whereas the element with the strongest gradient, [C/H], has the highest azimuthal scatter.
The trends for the overall scatter across $R = 0 - 20 \kpc$ reflect those of the radial gradient above.
Again, the trends in azimuthal and total variations are similar across these elements.

Finally, Figure~\ref{fig:RAD_and_VER_slope_and_AZIM_Fe_H_C_Mg_3plots_PUBLISH} (bottom) shows the vertical gradient, $\partial {\rm [X/H]} / \partial |Z|$ across $|Z| < 1 \kpc$, of all stars regardless of age, versus $R$, for [Fe/H], [C/H], and [Mg/H], using the same selection criteria as in Figure~\ref{fig:VER_slope_vs_R_PUBLISH}. Similar to the radial gradients for these elements, [Mg/H] has a shallower vertical gradient than [Fe/H], and [C/H] has a steeper vertical gradient than [Fe/H], both today and using coordinates at formation. But again, the trends with $R$, and the differences between today and formation, are similar.

Overall, while the exact slope of the gradients or the strength of the azimuthal variations do depend on the measured element, the overall trends, both today and at birth are similar, so we conclude that all of the trends that we explored for [Fe/H] hold true for other elemental abundances in FIRE-2.

\end{document}